\documentclass[11pt]{article}

\usepackage[T1]{fontenc}
\usepackage[utf8]{inputenc}
\usepackage[english]{babel}

\usepackage{graphicx}
\usepackage[a4paper,top=3.5cm,left=3cm,right=3cm,bottom=2.5cm]{geometry}

\usepackage{indentfirst}
\usepackage{caption}
\usepackage{subcaption}
\usepackage{float}
\usepackage{cancel}
\usepackage{xcolor}
\usepackage{lipsum}
\usepackage{comment}
\usepackage{hyperref}

\usepackage{amsmath}
\usepackage{amsfonts}
\usepackage{amssymb}
\usepackage{mathrsfs}
\usepackage{bm}
\usepackage{stackengine}
\usepackage{todonotes}

\begin{document}

\title{Integrated Information, a Complexity Measure for optimal  partitions}

\author{Otavio Citton \thanks{Current address: Bernoulli Institute for Mathematics, Computer Science and Artificial Intelligence, University of Groningen, Nijenborgh 9, 9747
AG Groningen, The Netherlands} \thanks{\href{mailto:o.c.citton@rug.nl}{o.c.citton@rug.nl}}  \, and Nestor Caticha \thanks{\href{mailto:ncaticha@usp.br}{ncaticha@usp.br}} \\
{\it Instituto de  Fisica, Universidade de  Sao Paulo}\\{ Sao Paulo, SP, Brazil}}

\maketitle

\begin{abstract}
    Motivated by the possible applications that a better understanding of consciousness might bring, we follow Tononi's idea and calculate analytically a complexity index for two systems of  Ising spins with parallel update dynamics, the homogeneous and a modular infinite range models. Using the information geometry formulation of integrated information theory,  we calculate the geometric integrated information index, $\phi_G(\Pi)$  for a fixed partition $\Pi$ with $K$ components and $\Phi=$max$_\Pi\phi_G(\Pi)$ for $K=2$ or $3$. For systems in the deep ferromagnetic phase, the optimal partition undergoes a transition  such that the smallest (largest) component is above (resp. below) its critical temperature. The effects of partitioning are taken into account by introducing site dilution. 
\end{abstract}


\section{Introduction}
There is no general agreement on what is meant by the term consciousness. 
The impossibility, so far, of physical explanations of the first person perception of pain or other forms of awareness,  may lead to either abandoning the physical description of consciousness or dismissing it as either an illusion or even non-existent, a folly of subjectivism. 
Taking as a given that the only certainty we can have are the inner personal experiences, Tononi's Integrated Information Theory (IIT) \cite{tononi2004} proposes to extract a quantitative signature $\Phi$, that would betray the existence of an inner complexity that conscious systems ought to have. 
As a working hypothesis,  ``time is what clocks measure'', may not be fully satisfying but permitted great progress and the refinement of ideas about time. Measuring or estimating IIT's $\Phi$ 
 has the potential of showing the direction of refining the question of what is it that science can say about consciousness.
It may be that ``consciousness is what $\Phi$ measures'' is just a stepping stone to better questions, experiments and predictions. 
Denying that a system lacks consciousness hinges on having a definition to assess whether the system has it or not.
There is a mountain to climb: from one side, obtaining consciousness as the emergent property of a physical system; from the opposite route, the analysis of how we can identify  whether  it is conscious from its necessarily complex intrinsic properties.
No matter how high these two routes meet, they will not convince those that seek a first person description of how we experience the world.

There is also no consensus on the mathematical definition of $\Phi$ and Tononi and collaborators and other groups have pressed on, presenting refinements and variations on the theme. In general, IIT proposes $\Phi$ as a measure of casual influences of the different parts, the constituents of the system. This idea stems from the postulated irreducibility of consciousness into some  sum over the parts. A physical system that satisfies these properties would be called a Physical Substrate of Consciousness (PSC).
See \cite{barbosa2020} and \cite{barbosa2021} for a description of the 
current state of the Desiderata of IIT, which is rather a declaration of what elements an acceptable theory ought to present, but not yet the basis for a unique tractable mathematical theory and hence many possibilities arise when trying to translate this declaration of purpose into something that can be calculated theoretically, measured experimentally and compared. The merit lies in presenting the first attempt to quantify and therefore enable  measuring consciousness.  It is certainly an unfinished chapter in the history of science.

Despite being incomplete in its logical structure,  which is usual in an incipient theory, several applications, see \cite{SarassoCasali2021} for an extensive review, point to the utility from a clinical perspective of being able to correlate clinical characterization to numerical estimates of complexity. 
For example, in \cite{casali2013} the Perturbational Complexity Index (PCI) related to IIT, is shown to serve as a quantitative characterization of 
the neural correlates of consciousness, a marker that could identify and measure the level of consciousness of comatose, anesthetized, sleeping or fully awake subjects.
A timid approach to the subject would claim that even if $\Phi$ has no relation to consciousness, it derives its importance from being a measure of the complexity of the dynamics.

The information geometry based framework to study complexity measures as ``distances'' between probability manifolds introduced by  Oizumi {\it et al.} \cite{oizumi2016} to embody IIT ideas, is central to this paper. The ``geometric integrated information'' $\Phi_G$ is the relative entropy of a full joint probability distribution relative to the product of marginals of disconnected subsets for a fixed partition. 
In all definitions of $\Phi$, a concept that is always present is that it is a measure of how different a system is to a partially disconnected version of itself, and the information geometry notion of distance (or divergence) seems natural to describe such a measure. Different measures of distance can be used, such as in  \cite{Oizumi2014}  and  \cite{Aguilera} which use the Wasserstein distance.
There are, however, technical difficulties in the calculation of either $\Phi$ or $\Phi_G$  which have limited such characterizations to theoretical models which may fail to qualify as complex. The first application to a system in the thermodynamic limit, by Aguilera and Di Paolo \cite{Aguilera}, deals with a kinetic Ising model to calculate a complexity index based on the Wasserstein metric, which yields results different from the geometric approach we follow here. 
 
In this paper we present a calculation of the geometric complexity index based on the ideas of Oizumi, Tsuchiya and Amari \cite{oizumi2016}. We study the two versions of the infinite range parallel update Ising model.
We can calculate the geometric complexity index for a general partition with  a given number of disconnected components, then $\Phi_G$ is defined as the maximum value, per spin, over all those partitions. We present results for partitions into two or three components for the homogeneous infinite range model. We also look at the model with a modular structure with two groups of spins, in a simple attempt to make an analogy with the separation of a brain into hemispheres. It has a natural partition into two components which is not necessarily the partition that maximizes integrated information.  In the thermodynamic limit of both models, $\Phi_G$ presents phase transitions.

A detailed comparison between some versions of $\Phi$ was made by Mediano \textit{et al.} \cite{mediano2019}. Although they have calculated those measures for a not so complex and small system, namely an autoregressive Gaussian model, it is still interesting to note the large variability of behaviors. Each definition of $\Phi$ seems to capture different aspects of the model, and the problem of choosing a single definition that allows us to quantify consciousness, if at all possible, remains open. This holds another motivation for studying different measures for a system in the thermodynamic limit, since the results may give insights on what it is measuring and if it has some property that would be interesting as a signature of consciousness.
\section{Geometric Integrated Information}
Consider a system with $N$ interacting classical Ising spins, indexed on a set $\Lambda \subset {\mathbb Z}$,   $|\Lambda| =N$. We denote by $X = \{ x_i |{i\in \Lambda}\}$ and $Y = \{ y_i|{i\in \Lambda}\}$ the variables that characterize the system's state at two consecutive steps of a discrete time dynamics. The spatio-temporal interactions between its elements are described by a distribution $P\left( X, Y\right)$\footnote{Conditional information about the details of the model is not shown for notational simplicity. }, the full model. Let  the distribution $Q\left( X,Y \right)$ be a member of a probability manifold $\cal M$,  defined by imposing some constraints, e.g. some of the interactions are disconnected. The difference between the full model and the disconnected model can be quantified by the  Kullback-Leibler divergence \cite{oizumi2016}:
\begin{equation}\label{eq: min kl divergence}
     D_{\text{KL}}\left[ P \| Q \right] = \sum_{X,Y} P\left( X, Y\right) \log \frac{P\left( X, Y\right)}{Q\left( X, Y\right)},
\end{equation}
a  measure of the strength of the influence in the complete system between the disconnected elements.
Let $\Pi$ be a partition of the set $\Lambda$ into non-overlapping subsets and denote by $I, J$ components of  partition $\Pi$. A variable $X_I=\{x_i | i\in I\}$ stands for the set of variables on a given component $I$ of $\Pi$.

Integrated information aims to quantify the amount of ``synergistic'' influences the whole system exerts on its future in excess of what the independent parts of the system do. Synergy is an unusual word in the context of Statistical Mechanics of phase transitions, and it probably refers to interactions between the system's degrees of freedom and their effect on emergent or collective properties. In the information geometry framework, this can be achieved by considering the following disconnection:
\begin{equation}\label{eq: disconection constraint}
    Q\left(Y_I | X\right) = Q\left(Y_I | X_I\right),
\end{equation}
for every subset $I\in \Pi$. This defines the manifold 
\begin{eqnarray}
    \mathcal{M}_G &:=& \Big\{Q(X,Y) \big| Q(Y_I|X) = Q(Y_I|X_I) , \forall I \in \Pi \Big\},\label{eq: manifold} 
    \end{eqnarray}
where the disconnected system lives.

The geometric integrated information \cite{oizumi2016} associated to a particular partition is then defined as 
\begin{equation}\label{eq: def integrated information}
    \phi_G(\Pi) = \min_{Q \in \mathcal{M}_G} \sum_{X,Y} P\left( X,Y \right) \log \frac{P\left( X,Y \right)}{Q\left( X,Y \right)}.
\end{equation}
This is a complexity index relative to the particular partition under consideration.

We start by showing a specific model calculation with the simplest partition,
 a bipartition $\Pi = \{I,J\}$ with non-overlapping  components $I\subset \Lambda$ and $J$, the complement of $I$ in $\Lambda$. We define $\mathcal{M}_{IJ}$ the probability manifold whose elements satisfy  the integrated information disconnection constraint \eqref{eq: disconection constraint} for this bipartition. Every distribution $Q_{IJ} \in \mathcal{M}_{IJ}$ must decompose as follows:
\begin{equation}\label{eq: geometric info factorization}
    Q_{IJ}\left(X,Y\right) = Q_{IJ}\left(X\right) Q_{IJ}\left(Y_I | X_I\right) Q_{IJ}\left(Y_J | X_J Y_I\right).
\end{equation}
The minimum of the KL divergence subject to such constraints is obtained  first, by supposing it exits and then, the minimization of the Lagrangian
\begin{align}
    \mathcal{L} = D_{\text{KL}}\left[ P \| Q_{IJ} \right] + \lambda \left( \sum_X Q_{IJ}\left(X\right) - 1 \right) + \sum_{X_I} \mu\left(X_I\right) \left( \sum_{Y_I} Q_{IJ}\left(Y_I | X_I\right) - 1 \right) +\nonumber\\
    + \sum_{X_J, Y_I} \nu\left(X_J, Y_I\right) \left( \sum_{Y_J} Q_{IJ}\left(Y_J | X_J Y_I\right) - 1 \right),
\end{align}
is solved by
\begin{eqnarray}
    Q_{IJ}^*\left(X\right) &=& P\left(X\right),\\
    Q_{IJ}^*\left(Y_I | X_I\right) &=& P\left(Y_I | X_I\right),\\
    Q_{IJ}^*\left(Y_J | X_J, Y_I\right) &=& P\left(Y_J | X_J, Y_I\right).
\end{eqnarray}
The integrated information $\phi_G(\Pi)$ for a bipartition $\Pi = \{I,J\}$ is:
\begin{equation}\label{eq: bipartition phiIJ}
    \phi_{IJ} = \sum_{X,Y} P\left(X,Y\right)\log\frac{P\left(Y|X\right) P\left(Y_I | X_J\right)}{P\left(Y | X_J\right) P\left(Y_I | X_I\right)},
\end{equation}
and the geometric integrated information per spin
\begin{equation}
    \Phi_G = \max_{\Pi}  \lim_{N\rightarrow \infty} \frac{1}{N}\phi_{IJ}.
\end{equation}
 
\section{Long range kinetic Ising Model}
To proceed we choose a particular version of the infinite range Ising model.
It is formulated with an intrinsic discrete dynamics and defined through the interaction at two consecutive times, hence it fits perfectly the information geometry description of the integrated information index.

The Hamiltonian is
\begin{equation}
    H\left(X,Y\right) = -  \sum_{i,j\in \Lambda}  J_{ij} x_j y_i,
\end{equation}
and on the assumption that it is conserved, maximum entropy leads to the distribution
\begin{equation}
    P\left(X,Y | {J}\right) = \frac{1}{Z} \exp\left\{ \beta  \sum_{i,j} J_{ij} x_j y_i \right\},
\end{equation}
where $\beta$ is an inverse temperature for the system and $Z$, the partition function, ensures normalization. At this point the interaction $J_{ij}$ is simply $J_0/N$, for the fully connected system, but may take other values in order to implement a disconnection.

From the product rule,  the transition probability distribution is
\begin{equation}
    P\left(Y | X; J\right) = \frac{P\left(X, Y | J\right)}{P\left(X| J\right)} = \frac{P\left(X, Y | J\right)}{\sum_Y P\left(X, Y| J\right)} = \frac{\exp\left\{\beta \sum_{i,j} J_{ij} x_j y_i\right\}}{\prod_i 2 \cosh\left(\beta \sum_{j} J_{ij} x_j\right)},
\end{equation}
a product of the transition distributions for each element,
\begin{equation}\label{eq: transition probability}
    P\left(Y | X; J\right) = \prod_{i=1}^{N} \frac{\exp\left\{ \beta h_i y_i \right\}}{2 \cosh\left( \beta h_i \right)} = \prod_{i=1}^{N} P\left(y_i | X; J\right),
\end{equation}
where the local field on site $i$ is     $h_i = \sum_{j=1}^{N} J_{ij} x_j$. 

We only consider the case with zero external field,   since $h\ne 0$ only diminishes complexity. The partition function is
\begin{equation}
    Z(\beta,J_0,N) = \sum_{X,Y}  \exp \left\{\beta \frac{J_0}{N} \sum_{ij}x_iy_j
    \right\}.
\end{equation}
Introducing a $\delta\left(\tilde m-\frac{1}{N}\sum_i x_i\right)$ $\delta\left(\tilde n-\frac{1}{N}\sum_i y_i\right)$ into the partition function:
\begin{equation}
    Z(\beta,J_0,N) = \int d\tilde m \frac{d\hat m}{2\pi}\int d\tilde n \frac{d\hat n}{2\pi} e^{\beta J_0N \tilde m\tilde n} e^{i\hat m\tilde m} 
    e^{i\hat n\tilde n} 
    e^{N\log {\mathcal{Z}}}.
\end{equation}
We have introduced the single site partition function
 $   \mathcal{Z} = \sum_{x,y} e^{-\mathcal{H}}$,
 the canonical partition function associated with the Hamiltonian $\mathcal{H}
= - \frac{i}{N}\left( \hat{m} x  + \hat{n} y\right)$.
The partition function can be written as 
\begin{eqnarray}
    Z(\beta,J_0,N) &=& \int  e^{-\beta N \mathcal{ F}(\tilde m,\tilde n;\beta,J_0)} d\tilde md\tilde n\\
    \beta F(\beta, J_0) &=& -\lim_{N\rightarrow \infty} \frac{1}{N}\log Z(\beta,J_0,N).
\end{eqnarray}
$F(\beta, J_0) $ is the free energy of the system per spin and we call
 $\mathcal{ F}(\tilde m,\tilde n;\beta, J_0)$ the free energy functional, with some abuse of language. Then
\begin{eqnarray}
F (\beta, J_0) &=& \stackunder{min}{$\tilde m,\tilde n$} \,\, \mathcal{ F}( \tilde m,\tilde n;\beta, J_0)\\
   \left( m (\beta, J_0,), n (\beta, J_0)\right) &=& \stackunder{argmin}{$\tilde m,\tilde n$} \,\, \mathcal{ F}(\tilde m,\tilde n;\beta, J_0).
\end{eqnarray}

For large $N$ the saddle point integration leads to 
\begin{equation}
    Z(\beta,J_0,N) = e^{\beta J_0N mn} e^{i\hat m m
    }e^{i\hat n  n 
    }e^{N\log {\mathcal{Z}}}
\end{equation}
where the values of $\tilde m, \tilde n $ and $\hat m, \hat n$ were substituted by the saddle point equation solutions.
By imposing that the derivative of the exponent with respect to each one of them is equal to zero,
we find:
\begin{eqnarray}
    \hat m &=& i N \beta J_0 n, \\
    \hat n &=& i N \beta J_0 m,
\end{eqnarray}
the saddle point values of $\hat m $ and $\hat n$ depend on the values of $n$ and $m$ respectively. Evaluating the $\hat m,\hat n$ integrals by constant phase integration:
\begin{eqnarray}
    \mathcal{H} &=& (\beta J_0 n   
    ) x  +(\beta J_0 m   
    ) y,\\
     \mathcal{Z} &=& \exp \left(\log 2\cosh( \beta J_0 n 
     )+\log 2\cosh( \beta J_0 m 
     )\right).
\end{eqnarray}
The saddle point equations for the expected values are simply 
\begin{eqnarray}
    m &=&  \tanh \left(\beta J_0 n \right), \\
    n &=&  \tanh \left(\beta J_0 m \right),
\end{eqnarray}
and their stable solutions  yield $m=n$. The free energy functional per spin is
\begin{eqnarray}
    \mathcal{F} (\tilde m,\tilde n;\beta, J_0) &=&  J_0 \tilde m\tilde n  -\frac{1}{\beta} \log 4\cosh( \beta J_0 \tilde n) \cosh( \beta J_0 \tilde m). \label{freeenergyfunctional}
\end{eqnarray}

When the order parameters are the solutions of the saddle point equations, the free energy functional is at is minimum and yields the thermodynamic free energy
\begin{equation}
   F(\beta) = J_0 m^2 - \frac{2}{\beta} \log (2\cosh(\beta J_0 m)). 
\end{equation}
The entropy of the system can be calculated directly from the Shannon expression evaluated at the saddle point solutions, or obtained using the thermodynamic relation, taking into account that $m(\beta)=\tanh \left(\beta J_0 m(\beta)\right)$
\begin{eqnarray}
    S[P(X,Y)]&=& - N\frac{\partial F}{\partial T} \\
    &=& 2N\left( - \beta J_0 m(\beta)^2 + \log 2\cosh(\beta J_0 m(\beta))\right),\label{entropy_full}
\end{eqnarray}
For $-1\leq \tilde m \leq 1$, define the function
\begin{eqnarray}
    \mathcal{S}(\tilde m,\beta, J_0, N)
    &=& 2N\left( - \beta J_0 \tilde m^2 + \log 2\cosh(\beta J_0 \tilde m)\right),\label{Scal}
\end{eqnarray}
such that when $\tilde m =m(\beta)$, it gives the value of the entropy of the full model $S(\beta,J_0,N):= S[P(X,Y)]$, but it is defined for any value of $\tilde m$. 

A similar calculation leads to the conditional entropy in equilibrium
   \begin{eqnarray}
    \frac{1}{\beta N}  S[P(Y|X)] &=& - J_0  m^2 + \frac{1}{\beta} \log 2\cosh(\beta J_0 m).\label{entropy_conditional}
\end{eqnarray}
From equations \eqref{entropy_full} and \eqref{entropy_conditional} we see that the entropy of the joint distribution is simply twice the entropy of the conditional on the past distribution.
This is interesting since  $S[P(X,Y)] = S[P(Y|X)] + S[P(X)]$, it follows that $S[P(X)] = S[P(Y|X)]$ and since the Hamiltonian is symmetric in  $X$ and $Y$, using $S[P(X,Y)] = S[P(X|Y)] + S[P(Y)]$, we conclude that in equilibrium $S[P(Y|X)] = S[P(Y)]$, which means that, for calculating the entropy,  the past is not informative about the future. Of course, the past is informative about the future before reaching equilibrium.

\section{Implementing disconnections by Site Dilution}

To evaluate $\phi_{IJ}$, expression \eqref{eq: bipartition phiIJ}, we need the disconnected transition probabilities distributions. We deal with this problem by introducing a method to calculate them: disconnect by site dilution. 
 Given a bipartition of $\Lambda$ into subsets $I,J$, introduce the set of dilution variables $\bm{\eta} = \{ \eta_i \}_{i\in \Lambda}$, with $\eta_i=1$ if $i\in I$ and $0$ if $i \in J$. 
There is a one to one correspondence between bipartitions and $\bm \eta$ configurations, so $\phi_{\bm\eta} = \phi_{IJ}$. 
The interaction matrix of the disconnected system is obtained  with the auxiliary $\bm \eta$ variables. For example, for an interaction matrix $J_{ij} \eta_i (1 - \eta_j)$, only when $i \in I$ and $j \in J$, $i$ and $j$ interact. 

The transition probabilities densities needed in equation  \eqref{eq: bipartition phiIJ} are:
\begin{eqnarray}
    P\left(Y | X ;J,\bm{\eta}\right) &=& \frac{\exp\left\{\beta \sum_{i,j}J_{ij} x_j y_i \right\}}{\prod_{i} 2\cosh\left(\beta \sum_{j} J_{ij} x_j\right)}, \label{eq: transition prob 1}\\
    P\left(Y_I | X_I;J,\bm{\eta}\right) &=& \frac{\exp\left\{\beta \sum_{i,j}J_{ij} \eta_i \eta_j x_j y_i \right\}}{\prod_{i} 2\cosh\left(\beta \sum_{j} J_{ij} \eta_i \eta_j x_j\right)} \label{eq: transition prob 2},\\
    P\left(Y_I | X_J;J,\bm{\eta}\right) &=& \frac{\exp\left\{\beta \sum_{i,j}J_{ij} \eta_i \left(1 - \eta_j\right) x_j y_i \right\}}{\prod_{i} 2\cosh\left(\beta \sum_{j} J_{ij} \eta_i \left(1 - \eta_j\right) x_j\right)},\\
    P\left(Y | X_J;J,\bm{\eta}\right) &=& \frac{\exp\left\{\beta \sum_{i,j}J_{ij} \left(1 - \eta_j\right) x_j y_i \right\}}{\prod_{i} 2\cosh\left(\beta \sum_{j} J_{ij} \left(1 - \eta_j\right) x_j\right)}.\label{eq: transition prob 4}
\end{eqnarray}
Our expression for $\phi_{\bm{\eta}}$ can be shown by  explicit calculation  to equal a difference of conditional entropies
\begin{equation}\label{eq: difference of conditional entropies}
    \phi_{\bm{\eta}} = S[P(Y_I | X_I)] + S[P(Y_J | X_J)] - S[P(Y|X)].
\end{equation}

This means that, for this type of simple model, the geometric $\phi_G$ is equal to the stochastic interaction complexity (see \cite{oizumi2016} and \cite{ito2020}), and can be interpreted as the amount of information gained when the components $I$ and $J$ are allowed to communicate with each other, in order to predict the future state. 
A word of warning about notation: $S[P(Y_I|X_I)]$ is not the entropy of an Ising model defined on component $I\subset \Lambda$, but rather of the distribution in expression \eqref{eq: transition prob 2}, which is defined in the whole $\Lambda$.

With $m$ the magnetization of the full system and $\gamma =\sum_{i \in \Lambda}\eta_i/N$, the fraction of spins in set $I$, 
the conditional entropies in equilibrium are
\begin{eqnarray}
    S[P(Y_I | X_I)] &=& -\beta J_0 \gamma^2 N m^2 + N\gamma \log 2\cosh(\beta J_0 \gamma m),\\
    S[P(Y_J | X_J)] &=& -\beta J_0 (1-\gamma)^2 N m^2  + N(1-\gamma) \log 2\cosh(\beta J_0 (1-\gamma) m,)\\
    S[P(Y|X)] &=& -\beta J_0 N m^2 + N \log 2\cosh(\beta J_0 m),
\end{eqnarray}
\begin{eqnarray}
    \frac{1}{N}  \phi_{\bm{\eta}} 
    &=& 2\beta J_0\gamma(1-\gamma)  m^2 \nonumber 
    +\log \frac{(\cosh^\gamma(\beta J_0 \gamma m)) (\cosh^{1-\gamma}(\beta J_0 (1-\gamma) m))}{\cosh(\beta J_0 m)}. 
\end{eqnarray}
It is important to stress that the value of $m$ is the magnetization of the $N$ units full model with interaction $J_0/N$ and
entropy given by $S(\beta;J_0,N)$.  Hence, using definition \eqref{Scal} we can write
\begin{eqnarray}
    \phi_{\bm{\eta}} &=& \frac{\gamma}{2}\mathcal{S}(m,\gamma\beta; J_0,N)+\frac{1- \gamma}{2} \mathcal{S}(m,(1-\gamma)\beta; J_0,N)-\frac{1}{2}S(\beta;J_0,N).
    \label{phi_entropy}
\end{eqnarray}
Due to the lack of spatial geometry of the original spin system, the influence of the partition is only through its size and not on the particular sets $(I,J)$. 
For a general partition $\Pi$ into $K$ non overlapping subsets of fractional sizes $ \{ \gamma_a\}$, with $\sum_{a=1}^K \gamma_a =1$
\begin{eqnarray}
    \phi_\Pi &=&\frac{1}{2} \sum_{a=1}^K \gamma_a \mathcal{S}(m,\gamma_a\beta; J_0,N) -\frac{1}{2} S(\beta;J_0,N). \label{phi_entropyK}
\end{eqnarray}

\subsection{Bipartition}
\begin{figure}
    \centering
\begin{subfigure}[t]{0.47\textwidth}
    \centering
    \includegraphics[width=\linewidth]{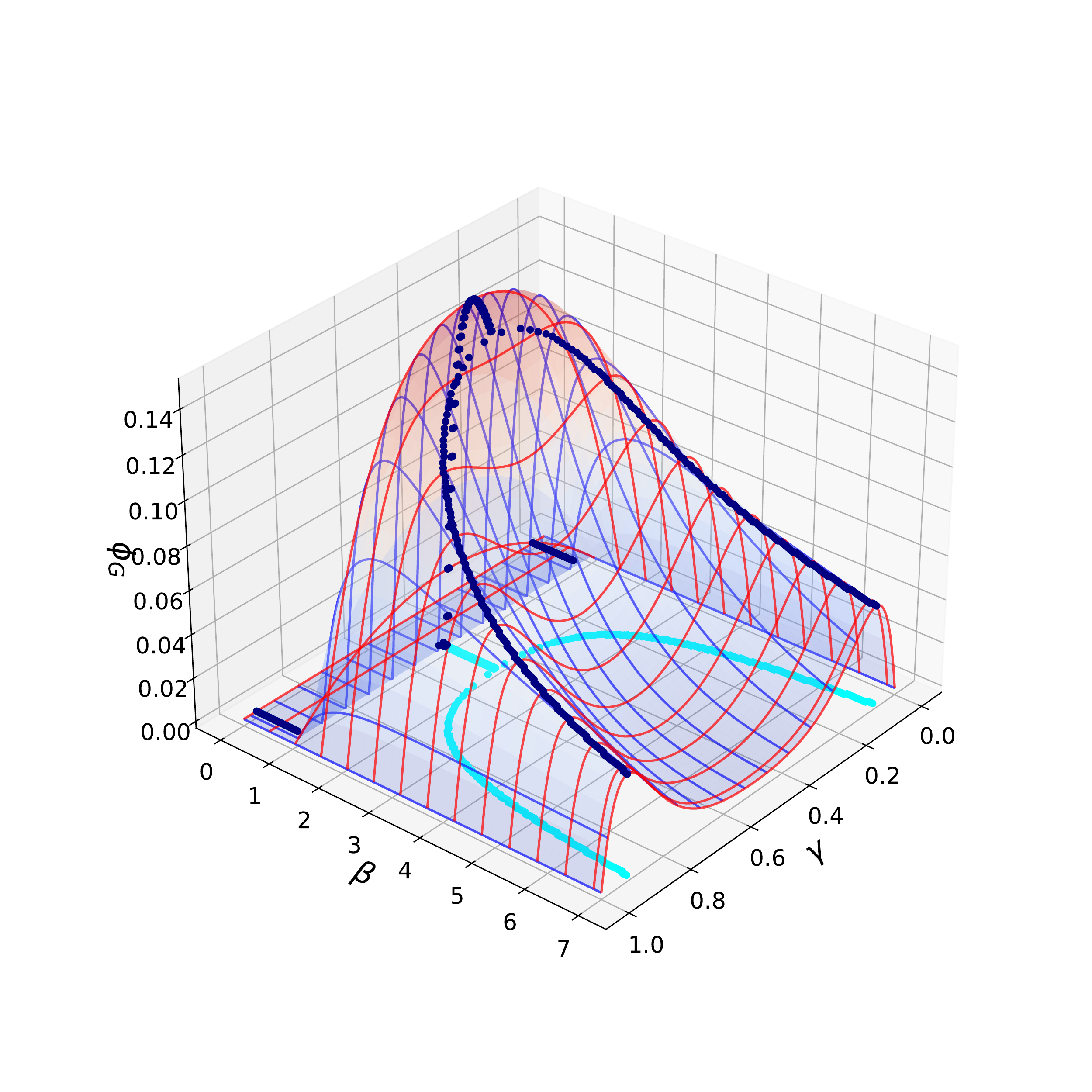}
    \caption{ }
    \label{fig:limitJ_zero}
\end{subfigure}
\hfill
\begin{subfigure}[t]{0.43\textwidth}
    \centering
    \includegraphics[width=\linewidth]{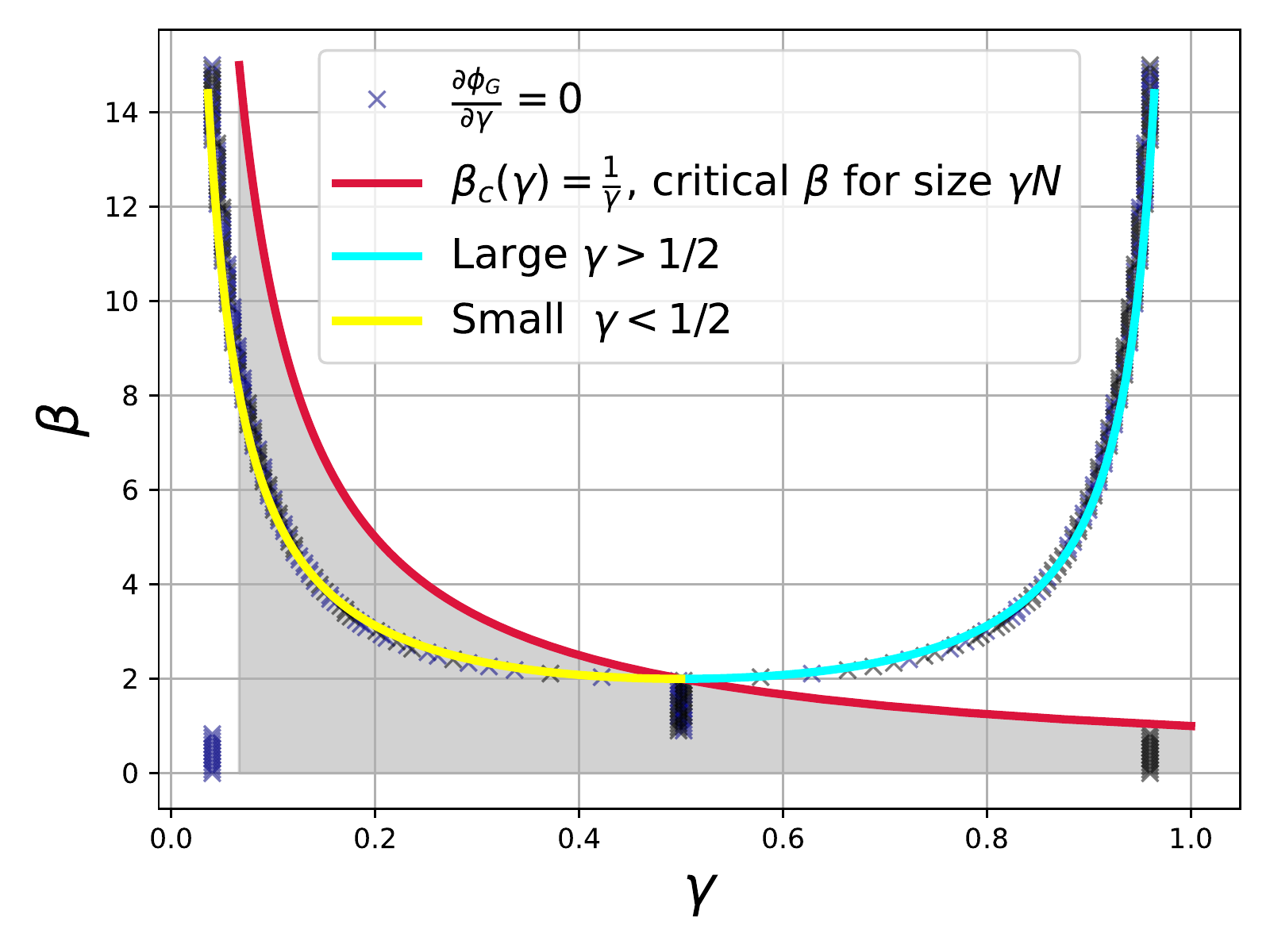}
    \caption{}
    \label{fig:optimalbif}
\end{subfigure}
\caption{Fig \ref{fig:limitJ_zero}: $\phi_{IJ}(\beta, \gamma)$. The dark circles show the maximum values of $\phi_{IJ}$ per spin for fixed $\beta$. The floor shows its projection into the $(\beta, \gamma)$ plane. 
Fig \ref{fig:optimalbif}: The  optimal bipartition in the  $(\beta, \gamma)$ plane. Symmetry breaks for $\beta> 2$, with the optimal partition, given by the inverse of $\beta_B(\gamma)$.  The small component remains in the (gray) paramagnetic region, below $\beta_c(\gamma)$ and the large component is in a ferromagnetic state, above $\beta_c(\gamma)$.}
\end{figure}
In figure \ref{fig:limitJ_zero} we show $\phi_{IJ}$ for a general bipartition. In the paramagnetic phase, the index is zero, which shows that the paramagnetic phase for the full system, is similar to that of the disconnected system. Statistically, the full model in the paramagnetic  phase behaves as if effectively disconnected. 
As soon as $\beta^{-1}$ becomes smaller than $\beta^{-1}_c = 1 $, the critical temperature of the full system measured in units of $J_0$, $\phi_{IJ}$ is different from zero and has a single maximum, at $\gamma = 1/2$. However, as $\beta$ goes above ${2}$, both disconnected systems with $\gamma = 1/2$ can present ferromagnetic order, and the value of $\gamma$ where   $\phi_{IJ}$ is maximum,  bifurcates away from $1/2$. This reflects the fact that the small component of the partition is paramagnetic and the large component is ferromagnetic. If both were ordered, the KL divergence to the full model would be smaller. At that temperature, both components can't be disordered. Hence, the disconnection  into equal sets  doesn't yield the largest integrated information loss to the full system. For a given $\beta$, the bipartition into $\gamma$ and $1-\gamma$ is optimal for (see figure \ref{fig:optimalbif}):
\begin{eqnarray}
\gamma &=& \frac{1}{2}+{\cal A}(\beta-\beta_c)^{\frac{1}{2}},\label{gammabif}
\end{eqnarray}
with values $\beta_c=2$ and  ${\cal A}\approx 0.400$ for  $\beta > 2$ and ${\cal A}=0$ for $\beta< 0$. Details are shown in \cite{SuppMat}.

\subsection{Tripartition}

\begin{figure}
    \centering
    \begin{subfigure}[t]{0.45\textwidth}\centering
        \includegraphics[width=\linewidth]{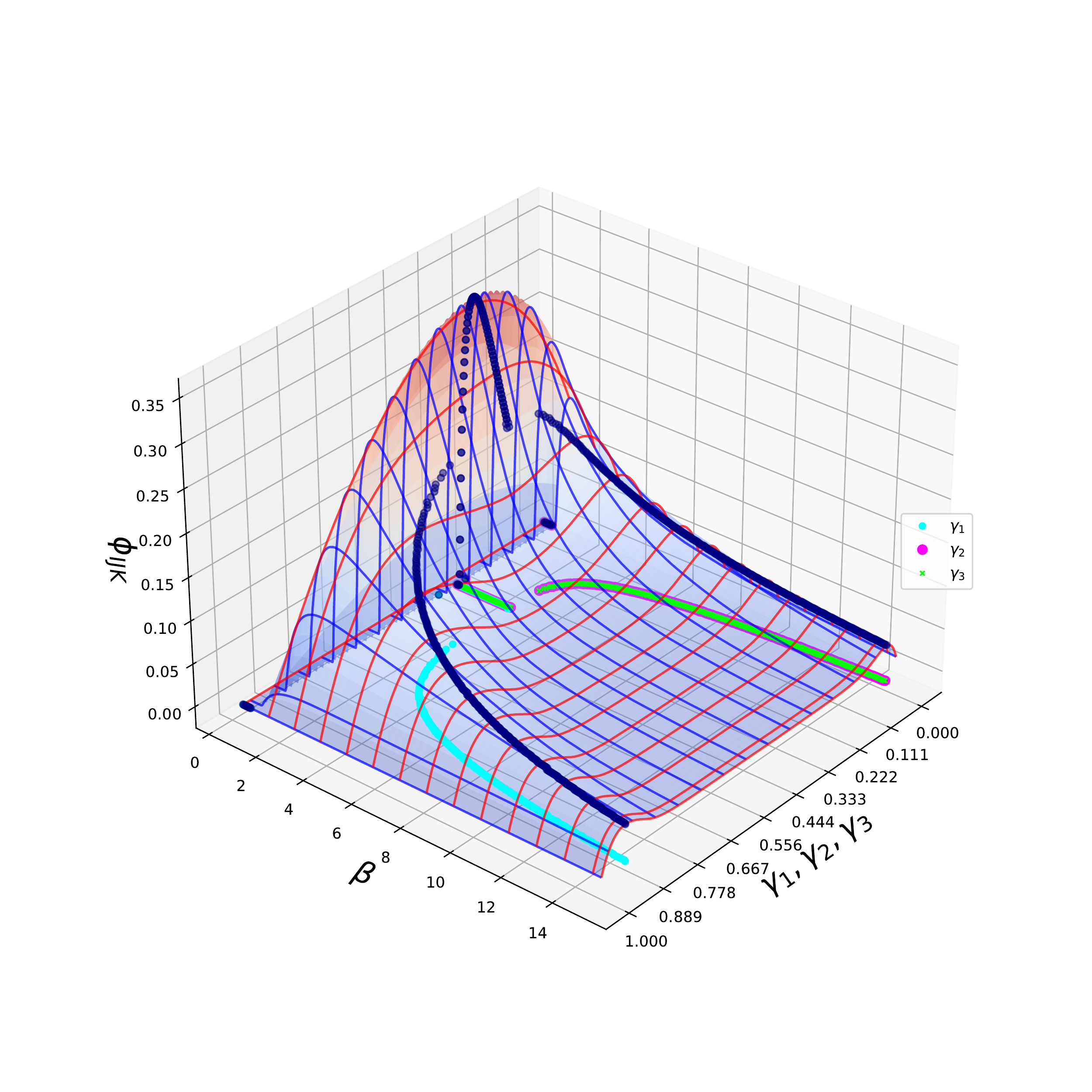}
        \caption{}
        \label{tripartition}
    \end{subfigure}
    \hfill
    \begin{subfigure}[t]{0.45\textwidth}\centering
        \includegraphics[width=\linewidth]{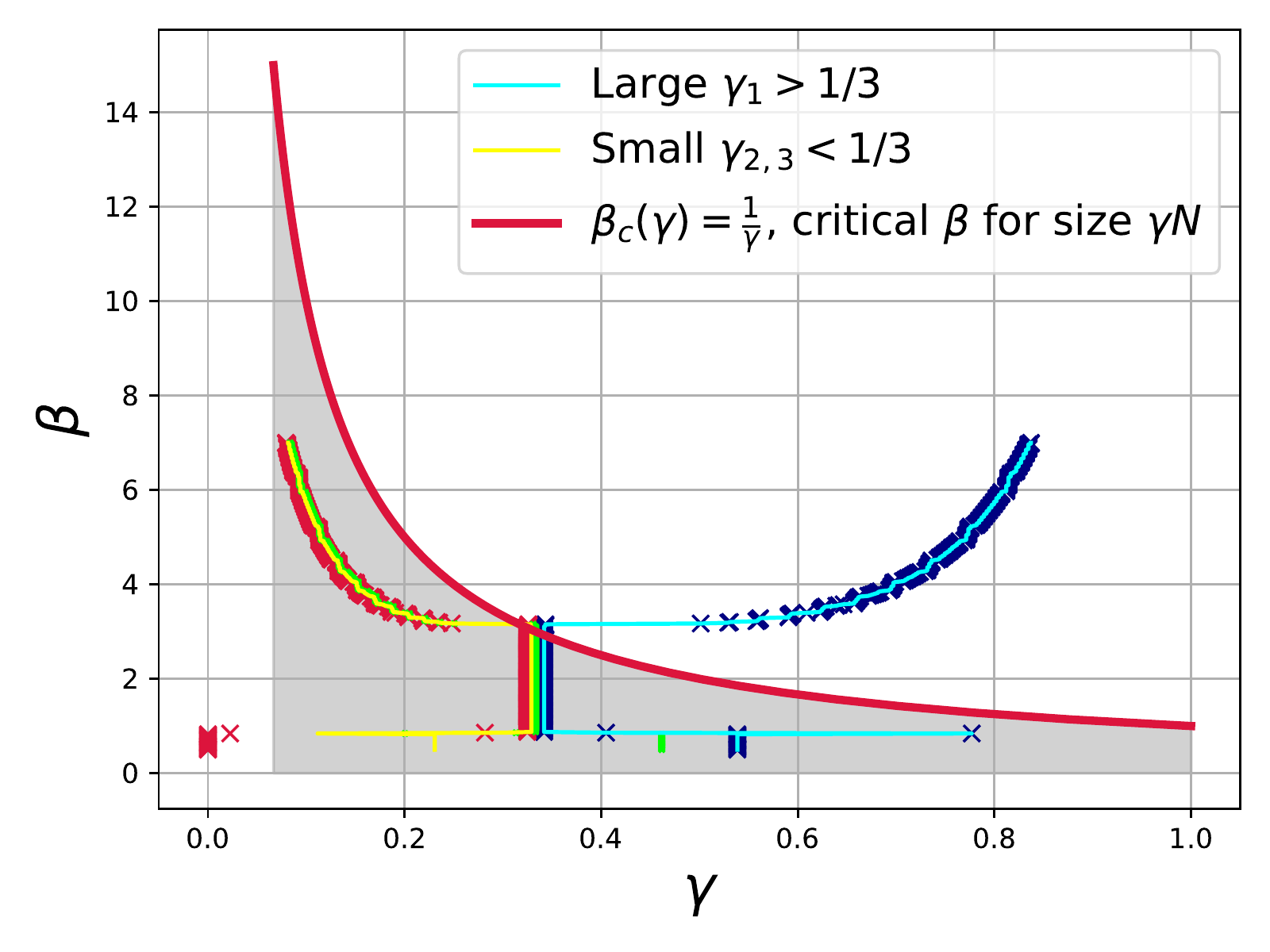}
        \caption{}
    \label{gamma_betatri}
    \end{subfigure}

    \caption{Fig \ref{tripartition}: $\phi_{IJK}(\gamma,\beta) =\text{max}_{\gamma_2}\phi_{IJK}(\gamma_1,\gamma_2,\gamma_3,\beta)$  for a partition into three subsets. The floor shows, as a function of $\beta$ the values of $\gamma_{1-3}$ that  yield the maximum value. Before the bifurcation at $\beta=3$, all $\gamma_i=1/3$. For larger $\beta$, one subset increases ($\gamma_1$) and the other two decrease, but remain equal. Fig \ref{gamma_betatri}: The  optimal tripartition in the  $(\beta, \gamma)$ plane. There is a discontinuous partially symmetry breaking transition at $\beta> 3$. The small components, with fractions less than $1/3$, remain in the (gray) paramagnetic region, below $\beta_c(\gamma)$ and the large component is in a ferromagnetic state, above $\beta_c(\gamma)$.} 
    \label{tripartitionB}
\end{figure}
Panel \ref{tripartitionB} shows $\phi_\Pi$ when $\Pi$ has three components: $\phi_{IJK}$. Call the fractional sizes $\gamma_a$, and choose their labels such that   $\gamma_1 \ge \gamma_2\ge \gamma_3$, with $\sum_i \gamma_i =1$. For a fixed $\gamma_a \ge \frac{1}{3}$, we calculate the value of $\gamma_b$ in the interval $[0,1-\gamma_a]$ that  maximizes $\phi_\Pi$ given by equation \ref{phi_entropyK}. Then $\gamma_1=$max$(\gamma_a,\gamma_b)$, $\gamma_2=$max$($min $(\gamma_a,\gamma_b), 1-\gamma_a-\gamma_b)$ and $\gamma_3=1-\gamma_1-\gamma_2$.
Again, in the paramagnetic region $\phi_{IJK}=0$. For $1 \le \beta \le 3$, the system is equi-partitioned $\gamma_1=\gamma_2=\gamma_3=1/3$. For  $\beta > 3$, a component with $\gamma_1 \ge 1/3$ can be ordered, leading to a symmetry breaking transition with $\gamma_2=\gamma_3= (1-\gamma_1)/2 < 1/3$.
 The symmetry is only partially broken since there is still a remnant symmetry, with the two small components equal in size. 

Partitioning the system into three subsets allows for a larger $\phi_\Pi$ than for partitions into two subsets. This is to be expected, since the KL divergences from a ferromagnetic state to a partitioned system increases with the size of the paramagnetic component of the partitioned system. It is expected, then that in general $\Phi_G$ grows with the number of components of $\Pi$. 
\section{Modular Ising model}
\begin{figure}
    \centering
    \includegraphics[width=.5\linewidth]{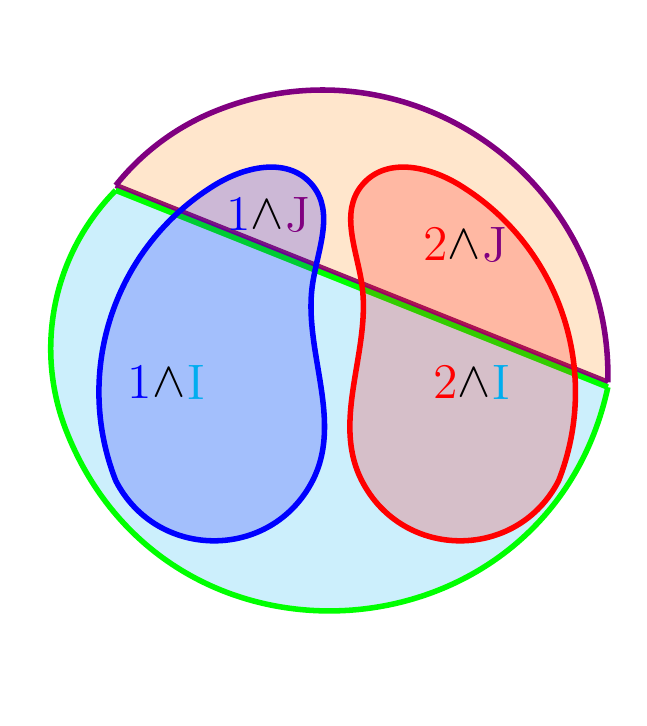}
    \caption{A system with a natural separation into modules $1$ and $2$ can be divided, for calculation of $\phi_{IJ}$ into arbitrary subsets $I$ and $J$, giving rise to four types of units.}
    \label{fig:conjmod}
\end{figure}

In the previous examples there is no natural partitioning of the system into components. It is possible that this index of complexity finds its utility when applied to neural systems which present a natural partition. For example, see figure \ref{fig:conjmod}, a system might be naturally divided into broad groups such as the hemispheres of a brain, or into a larger set of components, such as Brodmann areas. Of course, we are not ready to work with realistic models of a brain. We thus study the parallel updating version of a modular Ising model, similar to one studied by \cite{DomanySPC} in the context of super-paramagnetic clustering.

The system has $N$ Ising units, and again, the state of the system is given at time $t-1$ by $X = \left\{ x_i \right\}_{i=1}^{N}$, and  by $Y = \{ y_i\}_{i=1}^{N}$ at $t$. The Hamiltonian is:
\small
\begin{equation}
    H_M(X,Y) = - \frac{J_1}{N} \sum_{i,j=1}^{M} x_j y_i - \frac{J_2}{N} \sum_{i,j=M+1}^{N} x_j y_i - \frac{J_{21}}{N} \sum_{i=M+1}^{N}\sum_{j=1}^{M} x_j y_i - \frac{J_{12}}{N} \sum_{i=1}^{M}\sum_{j=M+1}^{N} x_j y_i.
\end{equation}
\normalsize
The first $M$ units belong to group $1$. They interact with their future through $J_1/N$ and with the future of the other $N-M$ units, belonging to group $2$, through $J_{21}/N$.  The $N-M$ elements interact with their future through $J_2/N$ and with group $1$ units future through $J_{12}/N$.
The equilibrium is described by a Gibbs distribution.
Call $\tilde \gamma_1 = M/N, \tilde \gamma_2=1-\tilde \gamma_1$, $\mathbf{m} = {m_1 \choose m_2}$,   $\mathbf{n} = {n_1 \choose n_2}$ and 
\begin{eqnarray}
    \mathbf{J} &=& \begin{bmatrix}
    J_{1}  & J_{12}  \\
    J_{21} & J_{2}
    \end{bmatrix}.
\end{eqnarray}
Using the same methods as in the previous section, we obtain for the free energy functional:

\begin{equation}
    {\mathcal{F}_M}(\mathbf{m,n};\beta,\mathbf{J}, \mathbf{\tilde \gamma}) 
    = \mathbf{n}^T \cdot \mathbf{J} \cdot \mathbf{m} - \frac{1}{\beta} \sum_{i=1,2}\tilde \gamma_i\log 4\cosh\left[\beta\left( \mathbf{n}^T \cdot\mathbf{J}\right)_i\right] \cosh\left[\beta\left( \mathbf{J} \cdot \mathbf{m}\right)_i\right],
\end{equation}
and the free energy

\begin{eqnarray}
    {{F}_M}(\beta,\mathbf{J}, \mathbf{\tilde \gamma}) &=& -\lim_{N \rightarrow \infty} \frac{1}{\beta N} \log Z,
    \end{eqnarray}
and the saddle point equations, for $i=1,2$:
\begin{eqnarray}
     n_i  &=&  \tilde \gamma_i \tanh\left[\beta (\mathbf{J}\cdot \mathbf{m})_i \right] \\
     m_i  &=&  \tilde\gamma_i \tanh\left[\beta(\mathbf{n}^T \cdot \mathbf{J})_i \right] \\
\end{eqnarray}
Again we obtain the entropy function from the derivative of the free-energy 
\begin{equation}
    {\mathcal{ S}_M}(\mathbf{m,n};\beta,\mathbf{J}, \mathbf{\tilde \gamma}) = -\beta \mathbf{n}^T \cdot \mathbf{J} \cdot \mathbf{m} + \sum_{i=1,2}\tilde \gamma_i\log 4\cosh\left[\beta\left( \mathbf{n}^T \cdot\mathbf{J}\right)_i\right] \cosh\left[\beta\left( \mathbf{J} \cdot \mathbf{m}\right)_i\right].
\end{equation}
While the modular Ising has a natural partition into groups $1$  and  $2$, the partition considered for the integrated information can, in principle, be different. We consider the partitions $I$ and $J$ of size $\gamma N$ and $(1-\gamma)N$ respectively. Introduce the mixing parameters: $\alpha$, the fraction of elements of module $1$ in partition $I$ and $\bar{\alpha} = \frac{\gamma - \alpha \tilde{\gamma}}{1 - \tilde{\gamma}}$,  the fraction of elements of module $2$ in partition  $I$. When $\gamma = \tilde\gamma=1/2$, the {\it natural} partition occurs for $\bar \alpha =1-\alpha$ and $\alpha=0$ or $1$. 

This  divides the units into four groups, of sizes:
\begin{equation}
    \begin{pmatrix}
        |1 \wedge I| & |2 \wedge I| \\
        |1 \wedge J| & |2 \wedge J|
    \end{pmatrix}
    =
    \begin{pmatrix}
        \alpha \tilde \gamma & \bar\alpha (1-\tilde \gamma) \\
        (1-\alpha)\tilde \gamma & (1-\bar \alpha)(1-\tilde \gamma) 
    \end{pmatrix}.
\end{equation}
The expression for $\phi_{\bm{{\eta}}}$, as shown in \cite{SuppMat} is
\begin{align}
    \phi_{\bm{\eta}}/N =& \beta\left( \mathbf{n}_J^T \cdot \mathbf{J} \cdot \mathbf{m}_I + \mathbf{n}_I^T \cdot \mathbf{J} \cdot \mathbf{m}_J \right) +\nonumber\\
    &- \tilde{\gamma} \log 2\cosh\left[\beta \left( \mathbf{J} \cdot \mathbf{m} \right)_1 \right] - (1 - \tilde{\gamma}) \log 2\cosh\left[\beta \left( \mathbf{J} \cdot \mathbf{m} \right)_2 \right] +\nonumber\\
    &+ \alpha \tilde\gamma \log 2\cosh\left[ \beta \left( \mathbf{J} \cdot \mathbf{m}_I \right)_1 \right] + \bar\alpha (1 - \tilde\gamma) \log 2\cosh\left[ \beta \left( \mathbf{J} \cdot \mathbf{m}_I \right)_2 \right] +\nonumber\\
    &+ (1 - \alpha) \tilde\gamma \log 2\cosh\left[ \beta \left( \mathbf{J} \cdot \mathbf{m}_J \right)_1 \right] + (1 - \bar\alpha) (1 - \tilde\gamma) \log 2\cosh\left[ \beta \left( \mathbf{J} \cdot \mathbf{m}_J \right)_2 \right]\label{phimodular}
\end{align}

The vectors $\mathbf{m}_I = {\alpha m_1 \choose \bar{\alpha} m_2}$ and $\mathbf{n}_I = {\alpha n_1 \choose \bar{\alpha} n_2}$ represent the partial magnetizations for the fraction of the system that belongs to the component $I$ and, analogously, $\mathbf{m}_J$ and $\mathbf{n}_J$ are the partial magnetizations for the elements that belong to the component $J$, such that $\mathbf{m} = \mathbf{m}_I + \mathbf{m}_J$ and $\mathbf{n} = \mathbf{n}_I + \mathbf{n}_J$. The measure of complexity $\phi_{\mathbf{\eta}}$ is not symmetric in the present and past variables, but with the approach to equilibrium their expected values become the same. 
\begin{figure}
    \centering
    \begin{subfigure}[t]{0.45\textwidth}
        \centering
        \includegraphics[width=\linewidth]{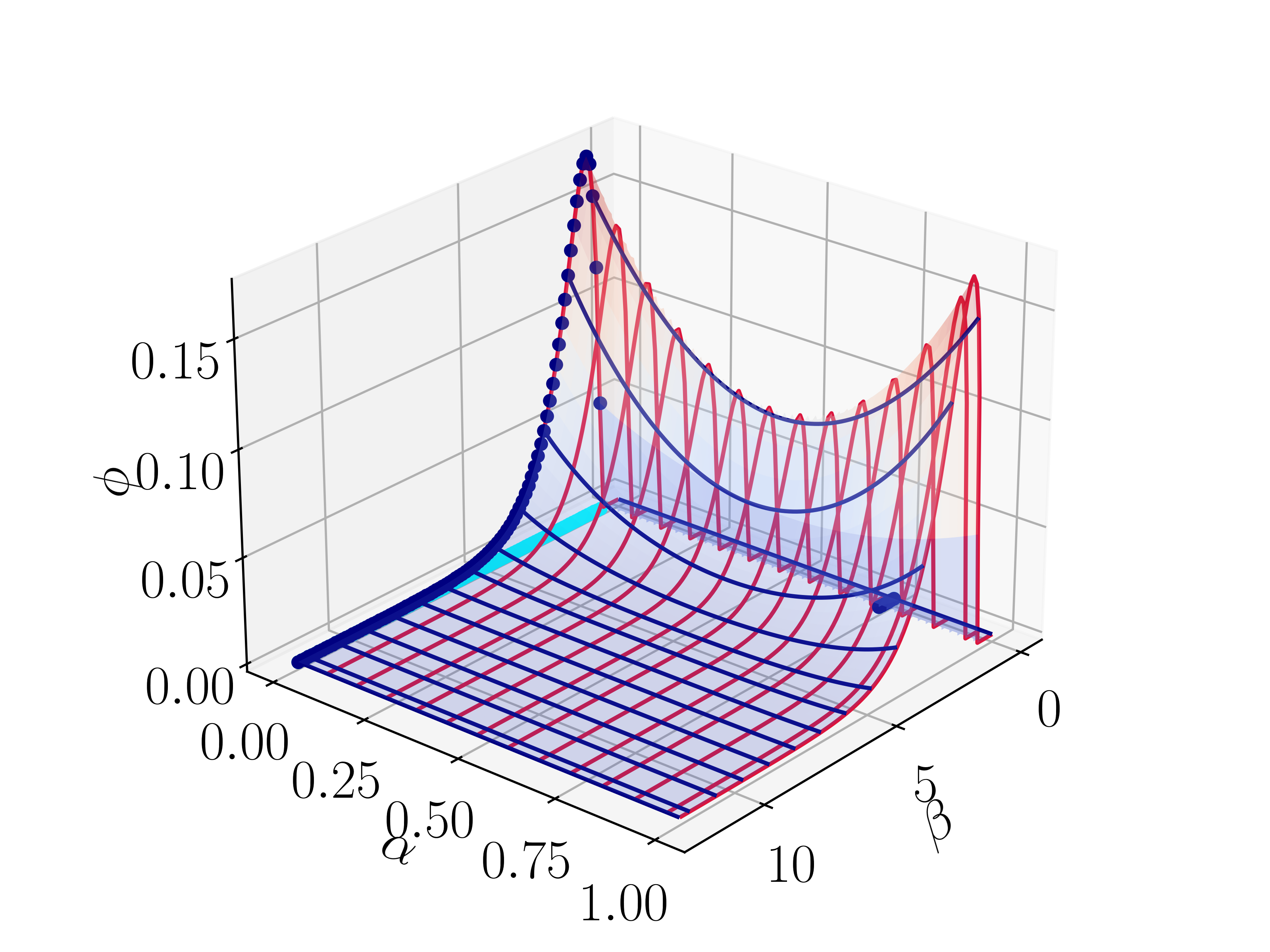}
        \caption{$J_1=J_2=1$, $J_{12}=J_{21}=2$. There is no  mixing for strong cross couplings. 
        Similar behavior for $\gamma=\tilde \gamma =1/2$ (shown) and for $1/4$ (not shown)
        } \label{phiModalphabetaA}
    \end{subfigure}
    \hfill
    \begin{subfigure}[t]{0.45\textwidth}
        \centering
        \includegraphics[width=\linewidth]{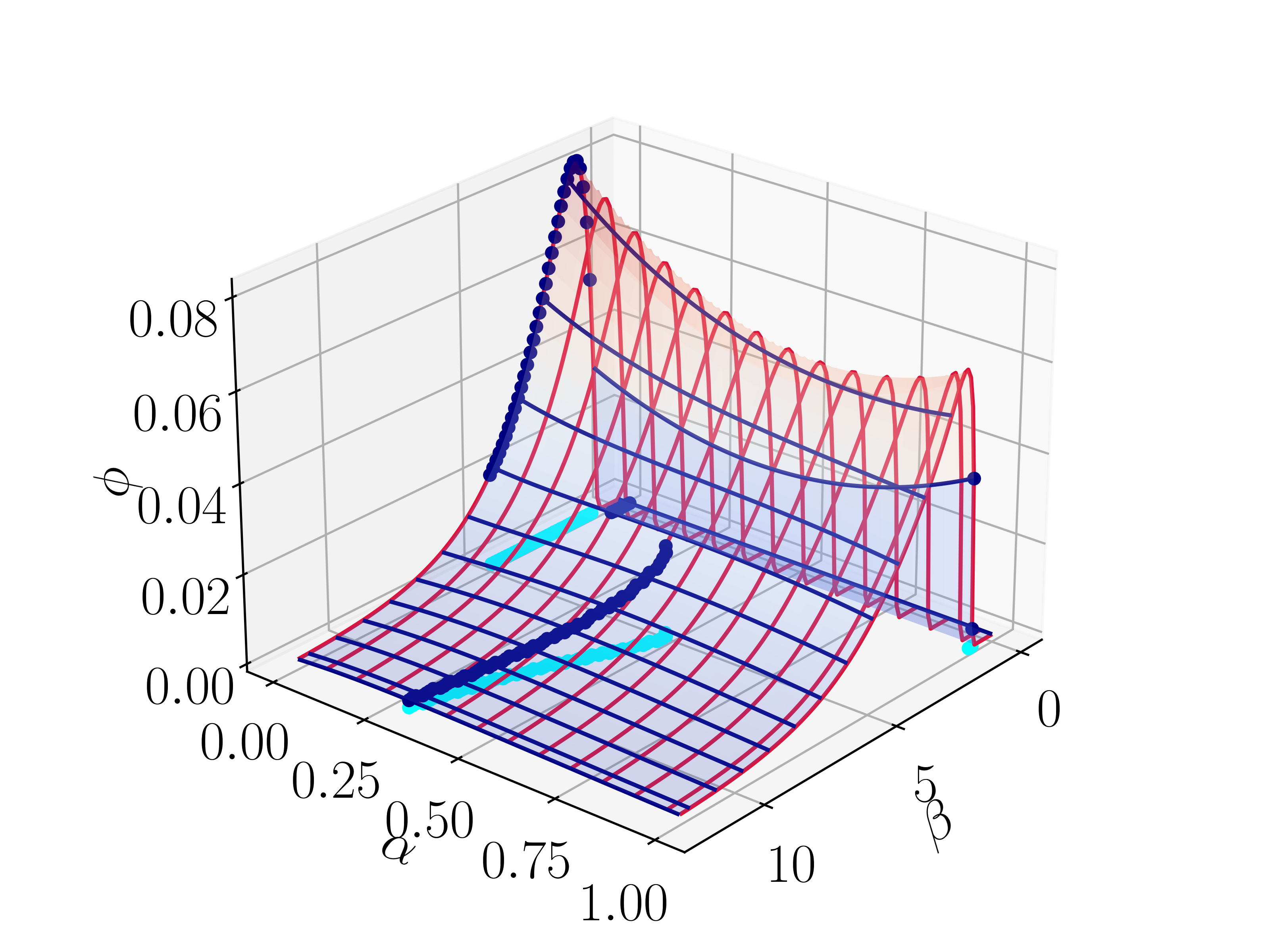}
        \caption{$2= J_{21} > J_1=J_2=1  >J_{12}= .5$. For asymmetric cross couplings and $\gamma=\tilde\gamma =1/4$ there is discontinuous transitions} 
        \label{phiModalphabetaB}
    \end{subfigure}

    \begin{subfigure}[t]{0.45\textwidth}
        \centering
        \includegraphics[width=\linewidth]       {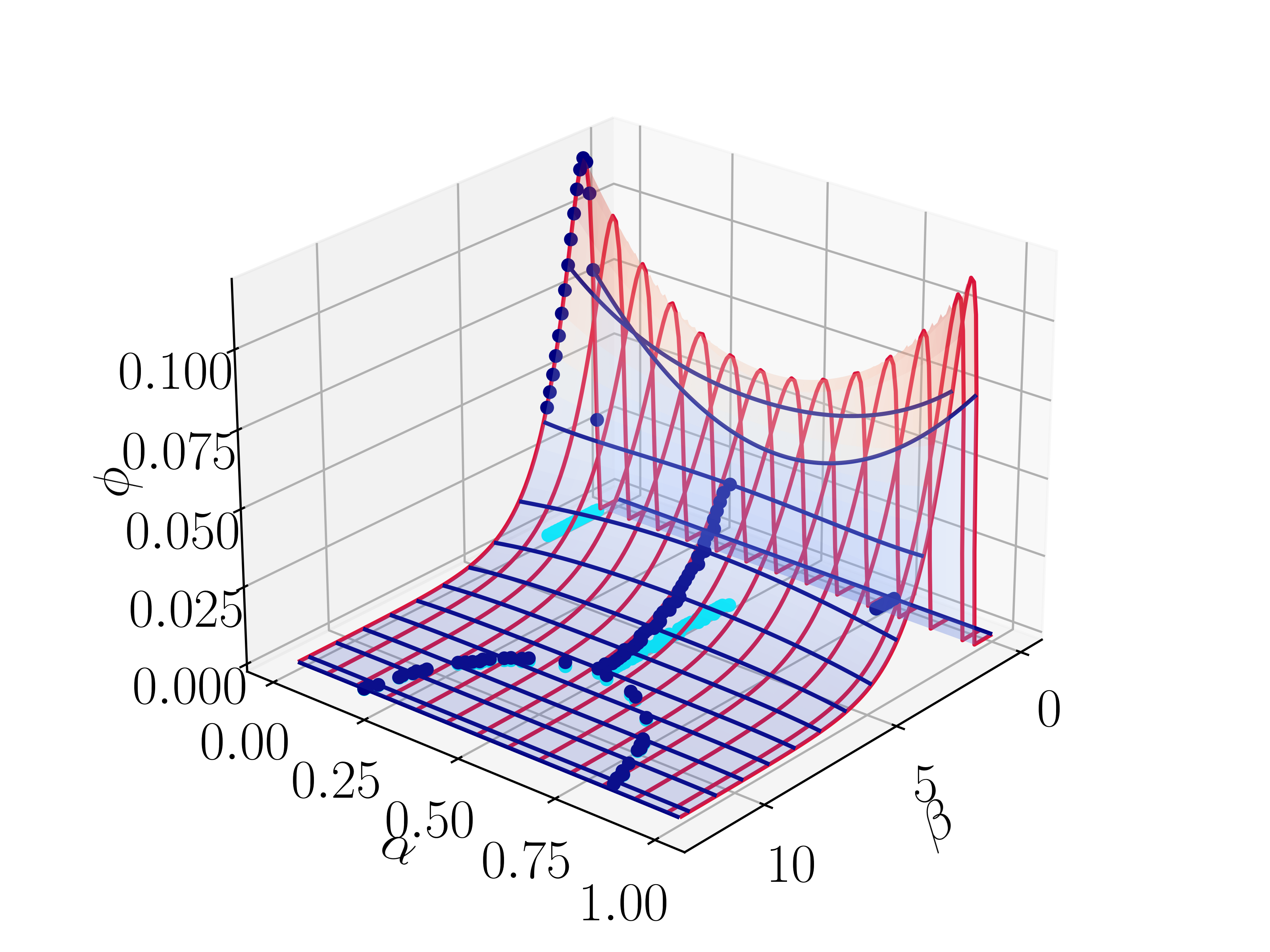}
        \caption{$2= J_{21} > J_1=J_2=1  >J_{12}= .5$. For asymmetric cross couplings and $\gamma=\tilde\gamma =1/2$ there are two transitions.} \label{phiModalphabetaC}
    \end{subfigure}
    \hfill
     \begin{subfigure}[t]{0.45\textwidth}
        \centering
        \includegraphics[width=\linewidth]{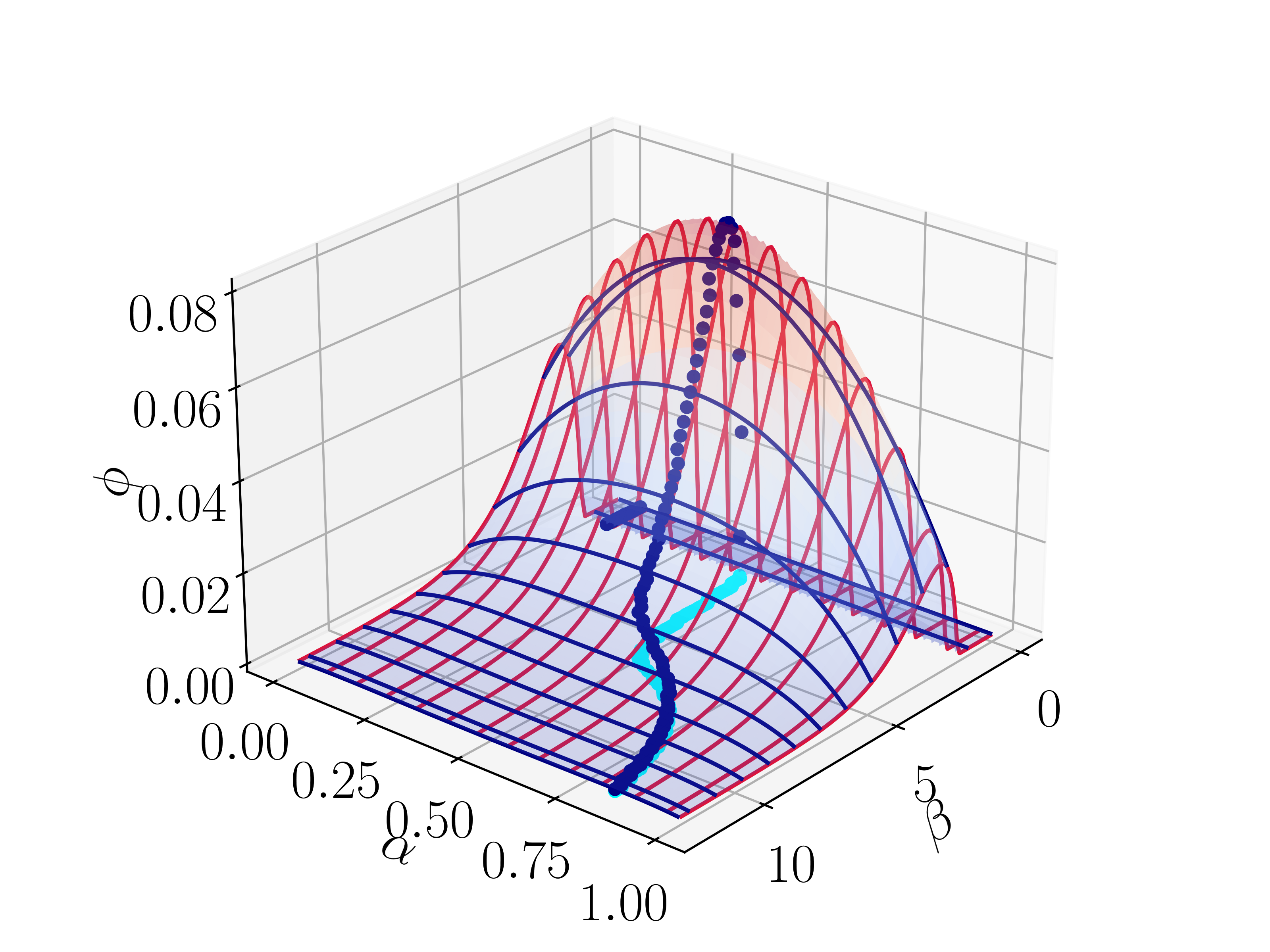}
        \caption{$J_1=J_2=1>J_{12}=J_{21}=0.4$. Weak cross couplings. Mixing symmetry breaking has a preference, $\gamma=\tilde \gamma =0.45$.} \label{phiModalphabetaD}
    \end{subfigure}
    \hfill
    \begin{subfigure}[t]{0.45\textwidth}
        \centering
        \includegraphics[width=\linewidth]{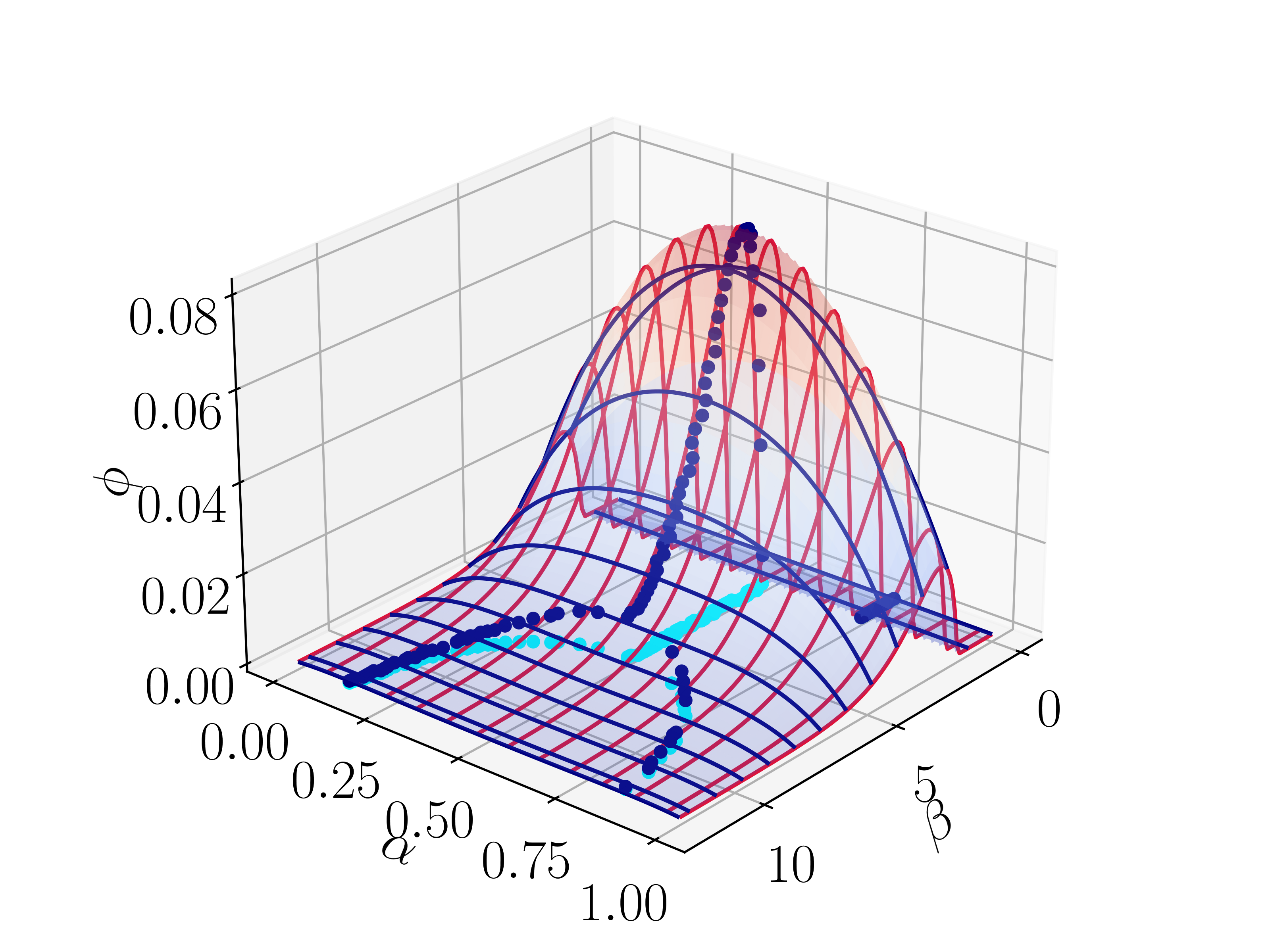}
        \caption{$J_1=J_2=1>J_{12}=J_{21}=0.4$. Weak cross couplings. Symmetry breaking in the mixing, $\gamma=\tilde \gamma =1/2$.} \label{phiModalphabetaE}        
    \end{subfigure}
    \hfill
    \begin{subfigure}[t]{0.45\textwidth}
        \centering
        \includegraphics[width=\linewidth]{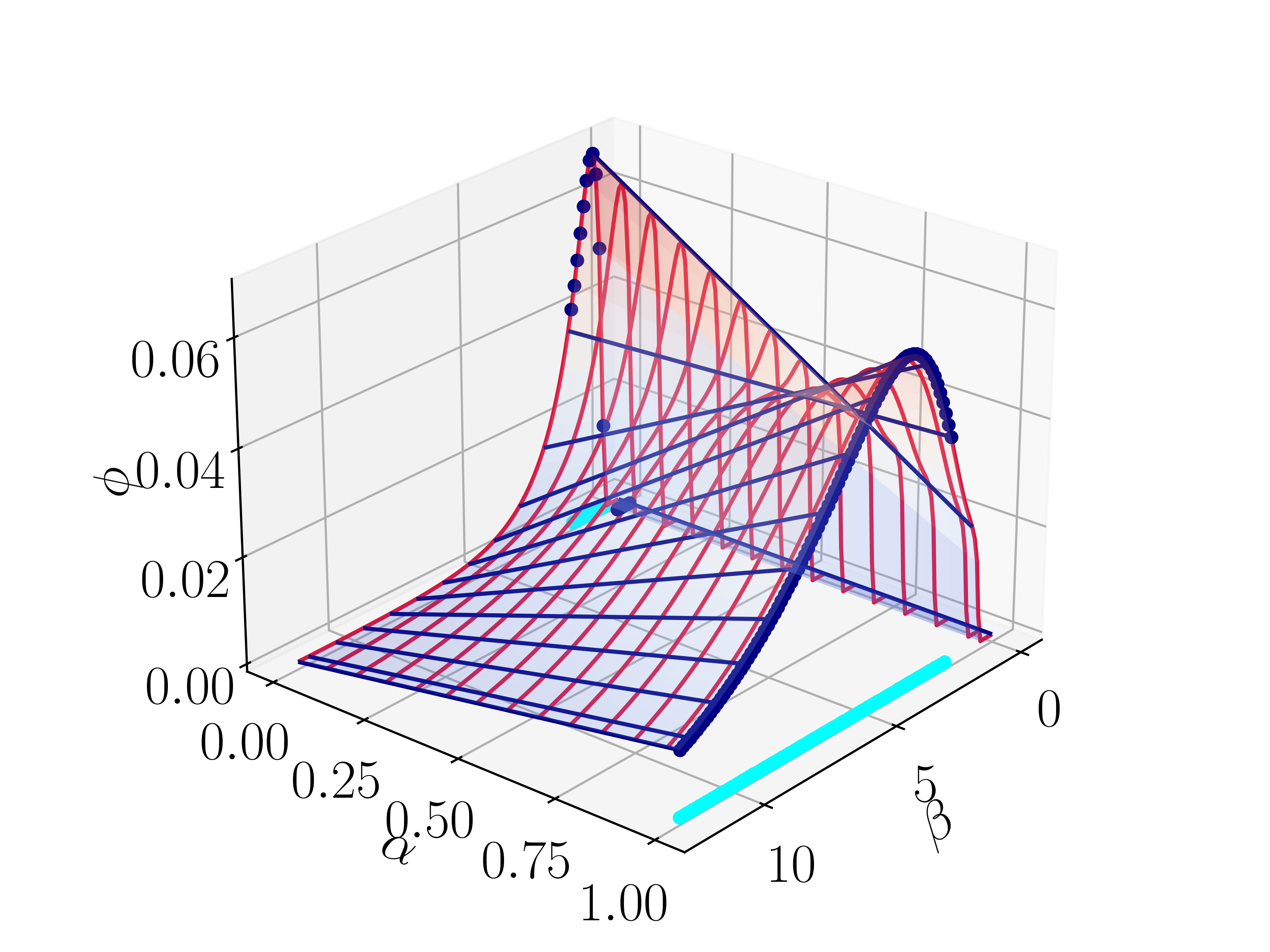}
        \caption{Connections that influence group $1$ are small and group $2$ are large: $J_1=J_{12}=.5$ and $J_2=J_{21}=2$. $\tilde \gamma =\gamma =1/4$} \label{phiModalphabetaF}
    \end{subfigure}
    \caption{ 
    The dark circles follow as a function of $\beta$, the maximum of $\phi$, with respect to  $\alpha$. The light circles in the floor show the projection of the maxima to the $(\alpha,\beta)$ plane. }
    \label{phiModalphabeta}
\end{figure}

 This expression is analyzed  in panel \ref{phiModalphabeta}, 
 which presents $\phi_{\bm{\eta}}$ as a function of the mixing parameter $\alpha$ and the inverse temperature $\beta$, for fixed  $\gamma$ and $\tilde \gamma$. We show results in \ref{phiModalphabetaA} to \ref{phiModalphabetaE}  when the intra-group connections are equal $J_1=J_2$ and the inter-group connection are either both higher $J_{12}, J_{21} > J_1$, both smaller $J_{12}, J_{21} < J_1$, or one smaller and one higher than the intra-group $J_{12}> J_1 >J_{21}$. 
 
 For large inter-group connections, the partition along the natural partition causes the largest difference, and thus there is no mixing: $\alpha=0$ or $1$ and  $\bar \alpha=1$ or $0$ for $\tilde \gamma= \gamma =1/2$, see \ref{phiModalphabetaA}.  All  members of one module are in the same partition. However, for $\tilde \gamma= \gamma =1/4$, which is not shown, there is a small and a large partition, as well as a small and a large module.  Then  $\alpha=0$ and $\bar \alpha=1/3$
  can't be zero and is  as small as possible. The inter-group connections are large and the partition that maximizes $\phi$ separates the modules as much as possible.
  
  Figure \ref{phiModalphabetaB} shows the case when the inter-group couplings take values larger and smaller than $J_1, J_2$ and the partitions are not equal in size, $\tilde \gamma =\tilde \gamma \neq 1/2$. Within the ferromagnetic region, but still low $\beta$, the partition separates the modules as much as possible as in the previous case. However, for sufficiently high $\beta$ there is a discontinuous  transition into a mixing state.  But when the partitions are symmetric, $\tilde \gamma =\tilde \gamma = 1/2$, see \ref{phiModalphabetaC}, in addition to the discontinuous transition to the $\alpha =\bar \alpha =1/2$ state,  there is a second continuous transition  into a $\alpha \neq 1/2$ state. The bifurcated curve shows the evolution of the values of $\alpha$ and $ \bar \alpha$. Again the reason for the bifurcation is that all the disconnected sets would enter the ferromagnetic phase, and the KL is reduced by preventing that, by reducing the size of two of the four subsets.

  For small inter-group connections the mixing for small $\beta>1$ goes to one half, for lower temperatures and then it changes. As $\beta$ increases for the asymmetric partitions $\tilde \gamma =\gamma \neq 1/2$, \ref{phiModalphabetaD},  changes into larger mixing of the small component. However, it bifurcates symmetrically when $\tilde \gamma =\gamma = 1/2$, \ref{phiModalphabetaE}, as in the case of the homogeneous system.
  
 Finally, \ref{phiModalphabetaF} shows the case $J_{12}= J_1 < J_2= J_{21}$, where the couplings to the future of group $1$ is smaller than the corresponding couplings to group $2$ and $\tilde \gamma =\gamma =1/4$. $\alpha$ changes abruptly from $0$ to $1$, while $\bar \alpha$ goes from $1/3$ to $0$ which is the natural partition. The transition occurs when the small partition $I$, containing only members of group $2$ becomes ferromagnetic. Then it is more disrupting to disconnect the inter-group couplings.

\section{Conclusions}
The main contribution of the IIT program is laying a road map to overcoming the barrier between a physical approach to a deduction of consciousness and the characterization of the degree to which an information processing system may be deemed conscious. While the first seems to be currently a vaguely posed problem, the second may lead to operational improvements of what it is meant by the many different aspects of consciousness, and even illuminating the first. Sadi Carnot motivated, specially by his father's failures to prove that no perpetuum mobile exists, just postulated their impossibility. The rewards were immense. Decades later, the probabilistic approach of Statistical Mechanics furnished a rational explanation for Carnot's leap, with the refinement of questions by the introduction of concepts like probability, entropy and later, the quantification of information. The study of complexity indices for different systems may lead to better questions about consciousness, and even if not,  the improvement of the computational techniques and understanding of complex systems is worthy of the efforts.

In this work, we have analyzed a complexity index, inspired by the geometrical approach in integrated information theory, for physical systems, in an attempt to better understand how it can be used to assess complexity. Our main experimental motivation was the application by Casali {\it et al.} \cite{casali2013} of related indices as markers of consciousness in patients whose brains are subject to different clinical conditions.
From a theoretical perspective, we follow the information geometric framework developed by Oizumi {\it et al.} \cite{oizumi2016}. By restricting the analysis to partitions into  $2$ and $3$ sets, we present the calculation of  the geometric integrated information index in two models.
An optimal partition is one in which the loss in information is greatest by removing the coupling between the components.  A phase transition in the size of the optimal partition was found. As the temperature is lowered in the ferromagnetic phase (below $T_c/2$ for bipartitions), symmetry is broken and an asymmetric bipartion becomes optimal. Going from the paramagnetic phase to the ferromagnetic phase $\phi$ increases. In the case of tripartitions, the symmetry is partially broken and the two small sets remain of the same size. 

We have addressed the question about what partition should be considered.  For the homogeneous model, it seems natural to consider the partition which at a given value of $\beta$, maximizes the KL divergence to the full system. However, interesting systems, serious candidates that deserve to be deemed conscious, are far from homogeneous. Information processing systems, arising from evolution, have a modular architecture of specialized macroscopic units and may suggest a particular partition as natural. There is a high degree of localization for different cognitive tasks in the brain, and it seems interesting to consider the partition of the system into something resembling the functional partition. There is no general theory to decide in this classification of different areas, in part because of the high connectivity among them. Thus, we look at the modular model which implements the idea of functional partitioning and explore the possibility of a different partitioning for IIT purposes, where the similarity of the functional and the IIT partition is measured by the mixing parameter $\alpha$.
 We defined $\Phi$ as the maximum over all partitions, while fixing $K$ the number of components of the partition. This invites the question of about its behavior as a function of the number of components;  the restricted scenario of $K=2,3$ supports that this may be monotonic in $K$.

It was shown, in \cite{oizumi2016}, that the manifold of interest for the calculation of $\Phi_G$, $\mathcal{M}_G$, contains a submanifold, $\mathcal{M}_S$, of the disconnected distributions whose components evolve independent of each other. The complexity measure calculated using the later manifold was called stochastic interaction, $SI$ \cite{ito2020},  and is always equal to the same difference of entropies we found in equation \eqref{eq: difference of conditional entropies}. We know from the hierarchical relation $\mathcal{M}_S \subset \mathcal{M}_G$, that $\phi_G \leq SI$. For the models analyzed here we have an equality, but the necessary conditions for this are still unclear, and an analysis of how the integrated information index and the stochastic interactions complexity measures differs from one another may yield interesting results.

The definition of $\Phi$ is dynamic and encompasses two consecutive states in time, so our equilibrium analysis only captures a small part of its phase space and more will be learned by studying the system out of equilibrium, since it is difficult to make the case for consciousness in a system in equilibrium.

\section*{Acknowledgments}
We thank Leonardo S. Barbosa, Marcus V. Baldo and O. Kinouchi for discussions. This work was partially supported by CAPES as part of project 88887.612147/2021-00. 

\bibliographystyle{unsrt}
\bibliography{references.bib}
\newpage
\appendix
\section{Supplementary Material} 
\subsection{$\gamma$ bifurcation}
Here we show equation \ref{gammabif}. With $m$ given by $m=\tanh \beta m$, write $0=\frac{1}{N} \frac{\partial \Phi}{\partial \gamma} =I_1+I_2+I_3 $ with
\begin{eqnarray}
I_1&=&2\beta(1-2\gamma)m^2 \\
I_2&=&\log \cosh{\beta \gamma m}- \log \cosh{\beta(1- \gamma) m}  \\
I_3&=&m\beta\left(\gamma \tanh (\beta \gamma m) -(1- \gamma) \tanh (\beta(1- \gamma) m)\right) 
\end{eqnarray}

The transition occurs near $\beta=2$ and $\gamma =\frac{1}{2}$, so define 
$\beta =2(1+\epsilon)$, $\gamma=(1+\theta)/2$, so that $1-\gamma =\frac{1-\theta}{2}$. Then $\beta \gamma = a+b$ and $\beta(1- \gamma)=a-b$, with $a=1+\epsilon \sim {\cal O} (1) $ and $b = \theta+\epsilon \theta \sim {\cal O} (\theta)$ is small near the transition. 
Substitution leads to 
\begin{eqnarray}
I_1&=&-4bm^2.
\end{eqnarray}
Write for short $t_a =\tanh(am)$ and $t_b=\tanh(bm)$ and expand $t_b$: $\tanh(x) = x- x^3/3$, and for $b\neq 0$
\begin{eqnarray}
I_2
&\approx&  2t_a b m  -\frac{2m^3b^3}{3}t_a(1-t_a^2) +{\cal O}(b^4)
\end{eqnarray}

\begin{eqnarray}
I_3 &\approx& 2m\left(t_a+am(1-t_a^2)\right)b -2m^3(1-t_a^2)\left(t_a -amt_a^2+ \frac{1}{3}a m\right)b^3 +{\cal O}(b^4)
\end{eqnarray}

The maximum with $\gamma$:
\begin{align}
    0 = \frac{1}{N} \frac{\partial \Phi}{\partial \gamma} =& \left(-4 m^2+2t_a  m + 2m\left(t_a+am(1-t_a^2)\right)\right)b +\nonumber\\
    &- 2m^3(1-t_a^2)\left(\frac{4}{3}t_a -amt_a^2+ \frac{1}{3}a m  \right)b^3.
\end{align}
For $b= a\theta=\frac{\beta}{2}\theta \neq 0$, $\theta= 2(\gamma- 1/2)$
\begin{eqnarray}
    \gamma &=&\frac{1}{2}+{\cal  A} \left(\beta -\beta_c\right)^{\frac{1}{2}}\label{bifurcaexato}
\end{eqnarray}
where
\begin{eqnarray}
\beta_c &=& \frac{4(m-t_a)}{m(1-t_a^2)} =2 \label{betac}\\
{\cal A}&=& \left(\frac{1}{2\beta^2} \frac{ 1}{\left(\frac{4}{3}t_a -amt_a^2+ \frac{1}{3}a m \right)}\right)^\frac{1}{2}
\end{eqnarray}

To prove \ref{betac} use that $m=\tanh \beta m$ is the magnetization of the full system, and $\beta =2(1+\epsilon)=2a $,  then $t_a =  \tanh(am) = \tanh(\beta m/2)$
\begin{eqnarray}
     \tanh(\beta m) = m&=& \frac{2t_a}{1+t_a^2} = t_a + t_a\frac{1-t_a^2}{1+t_a^2}\\
     \frac{4(m-t_a)}{(1-t_a^2)} &=& \frac{4t_a}{1+t_a^2} =2m
\end{eqnarray}
proving \ref{betac}. Numerically, at $\beta=2$, ${\cal A}= 0.40000$.

\subsection{ $\phi_\eta$ for the modular Ising model}

Here are presented the steps that lead to equation \ref{phimodular}, $\phi_{\bm{\eta}}$ for the Modular Ising Model. Writing $\phi_{\bm\eta}$ as a difference of conditional entropies explicitly
\begin{equation}
    \phi_{\bm{\eta}} = \sum_{X,Y} P\left(X,Y | \mathbf{J}\right) \left(\beta \sum_{i=1}^{N} y_i h_{i|D} - \sum_{i=1}^{N} \log 2 \cosh(\beta h_i) + \sum_{i=1}^{N} \log 2 \cosh(\beta h_{i|S}) \right),
\end{equation}
where we introduce the quantities $h_{i|S}$ and its complementary $h_{i|D}$, defined as
\begin{eqnarray}
    h_{i|S} &=& \left\{
    \begin{array}{l l}
    \sum_j J_{ij} \eta_j x_j, & \text{for} \quad i \in I\\
    \\
    \sum_j J_{ij} \left(1 - \eta_j\right) x_j, & \text{for} \quad i \in J
    \end{array}
    \right.,\\
    h_{i|D} &=& \left\{
    \begin{array}{l l}
    \sum_j J_{ij} \left(1 - \eta_j\right) x_j, & \text{for} \quad i \in I\\
    \\
    \sum_j J_{ij} \eta_j x_j, & \text{for} \quad i \in J
    \end{array}
    \right.,
\end{eqnarray}
as the fields generated on $i$ by the elements in the \textit{same} partition and in the \textit{different} partition, respectively. We immediately note that the total field is $h_i = h_{i|S} + h_{i|D}$.
Consider the 3 averages separately and call then $\phi_A$, $\phi_B$ and $\phi_C$, given by
\begin{eqnarray}
    \phi_A &=& \beta \sum_{X,Y} P\left(X,Y | \mathbf{J}\right) \sum_{i=1}^{N} y_i h_{i|D},\\
    \phi_B &=& - \sum_{X,Y} P\left(X,Y | \mathbf{J}\right) \sum_{i=1}^{N} \log 2 \cosh(\beta h_i),\\
    \phi_C &=& \sum_{X,Y} P\left(X,Y | \mathbf{J}\right) \sum_{i=1}^{N} \log 2 \cosh(\beta h_{i|S}).
\end{eqnarray}

The procedure we follow for the calculation of these averages is to consider auxiliary distributions. The distributions we are interested are those that are parameterized in such a way that the original distribution for the modular Ising model, can be recovered by a particular choice of the parameters. To calculate $\phi_A$, $\phi_B$ and $\phi_C$, consider, respectively,
\small
\begin{eqnarray}
    P_A\left(X, Y|\mathbf{J}; \bm\eta, \beta_S, \beta_D\right) &=& \frac{1}{Z_A} \exp\left\{ \sum_i \left( \beta_S h_{i|S} + \beta_D h_{i|D} \right)y_i \right\},\\
    P_B\left( X | \mathbf{J}; \bm\eta, \beta, \lambda \right) &=& \frac{1}{Z_B} \exp\left\{ \lambda \sum_i \log2\cosh\left( \beta \sum_j J_{ij} x_j \right) \right\},\\
    P_C (X,Y | \mathbf{J}; \bm\eta, \beta, \lambda_S) &=& \frac{1}{Z_C} \exp\left\{ \beta\sum_{i,j} J_{ij} x_j y_i + \lambda_S \sum_i \log 2\cosh\left( \beta \sum_j J_{ij} \theta_{ij}^{S} x_j \right) \right\}.
\end{eqnarray}
\normalsize
We note that by setting $\beta_S = \beta_D = \beta$, $\lambda=1$ and $\lambda_S = 0$ they all become the equilibrium distribution for the modular Ising model\footnote{With exception of $P_B$ which, for $\lambda = 0$, becomes the marginalized distribution $P(X|\mathbf{J}) = \sum_Y P(X,Y |\mathbf{J})$, which is not a problem, since $\phi_B$ only depends on the variables $X$.}, and the desired averages can be easily calculated as derivatives of the free energies:
\begin{eqnarray}
    \phi_A &=& \beta \left. \frac{\partial}{\partial \beta_D} \log Z_A \right|_{\beta_S = \beta_D = \beta},\\
    \phi_B &=& - \left. \frac{\partial}{\partial \lambda} \log Z_B \right|_{\lambda=1},\\
    \phi_C &=& \left. \frac{\partial}{\partial \lambda_S} \log Z_C \right|_{\lambda_S = 0}.
\end{eqnarray}

\subsubsection{$\log Z_A$ calculation}

\begin{equation}
    Z_A = \sum_{X,Y} \exp\left\{ \sum_{i,j} J_{ij} x_j y_i \left( \beta_S \theta_{ij}^S + \beta_D \theta_{ij}^D \right)\right\},
\end{equation}
we introduce here the projectors $\theta_{ij}^S = \eta_i \eta_j + (1-\eta_i)(1-\eta_j)$ and $\theta_{ij}^D = \eta_i(1-\eta_j) + (1-\eta_i)\eta_j$, that take into account the cases where $i$ and $j$ belongs to the same component and to different components, respectively.

Introducing integrals over Dirac delta distributions and using its Fourier representation
\small
\begin{align}
    Z_A =& \int\frac{dm_1^I d\hat{m}_1^I}{2\pi/N}\int\frac{dm_1^J d\hat{m}_1^J}{2\pi/N} \int\frac{dm_2^I d\hat{m}_2^I}{2\pi/N}\int\frac{dm_2^J d\hat{m}_2^J}{2\pi/N} \int\frac{dn_1^I d\hat{n}_1^I}{2\pi/N}\int\frac{dn_1^J d\hat{n}_1^J}{2\pi/N} \int\frac{dn_2^I d\hat{n}_2^I}{2\pi/N}\int\frac{dn_2^J d\hat{n}_2^J}{2\pi/N} \times\nonumber\\
    &\times \exp\bigg\{ \sum_{i=1}^{M} \log\zeta_i^{(1)} + \sum_{i=M+1}^{N} \log\zeta_i^{(2)} + \beta_S J_1 N \left( m_1^I n_1^I + m_1^J n_1^J \right) + \beta_D J_1 N \left( m_1^I n_1^J + m_1^J n_1^I \right) + \nonumber\\
    &+ \beta_S J_2 N \left( m_2^I n_2^I + m_2^J n_2^J \right) + \beta_D J_2 N \left( m_2^I n_2^J + m_2^J n_2^I \right) + \beta_S J_{12} N \left( m_2^I n_1^I + m_2^J n_1^J \right) +\nonumber\\
    &+ \beta_D J_{12} N \left( m_2^I n_1^J + m_2^J n_1^I \right) + \beta_S J_{21} N \left( m_1^I n_2^I + m_1^J n_2^J \right) + \beta_D J_{21} N \left( m_1^I n_2^J + m_1^J n_2^I \right) +\nonumber\\
    &+ iN\left( m_1^I\hat{m}_1^I + m_1^J\hat{m}_1^J + m_2^I\hat{m}_2^I + m_2^J\hat{m}_2^J + n_1^I\hat{n}_1^I + n_1^J\hat{n}_1^J + n_2^I\hat{n}_2^I + n_2^J\hat{n}_2^J \right)  \bigg\},
\end{align}
\normalsize
where
\begin{eqnarray}
    \zeta_i^{(1)} &=& \sum_{x_i, y_i} e^{-i\left( \hat{m}_1^I \eta_i + \hat{m}_1^J (1-\eta_i) \right) x_i -i\left( \hat{n}_1^I \eta_i + \hat{n}_1^J (1-\eta_i) \right) y_i},\\
    \zeta_i^{(2)} &=& \sum_{x_i, y_i} e^{-i\left( \hat{m}_2^I \eta_i + \hat{m}_2^J (1-\eta_i) \right) x_i -i\left( \hat{n}_2^I \eta_i + \hat{n}_2^J (1-\eta_i) \right) y_i}.
\end{eqnarray}

Note that the integrand is an exponential with exponent proportional to $N$, thus we can use the saddle point integration method and obtain
\small
\begin{align}
    \log Z_A =& -\beta_S J_1 N \left( m_1^I n_1^I + m_1^J n_1^J \right) - \beta_D J_1 N \left( m_1^I n_1^J + m_1^J n_1^I \right) - \beta_S J_2 N \left( m_2^I n_2^I + m_2^J n_2^J \right) +\nonumber\\
    &- \beta_D J_2 N \left( m_2^I n_2^J + m_2^J n_2^I \right) - \beta_S J_{12} N \left( m_2^I n_1^I + m_2^J n_1^J \right) - \beta_D J_{12} N \left( m_2^I n_1^J + m_2^J n_1^I \right) +\nonumber\\
    &- \beta_S J_{21} N \left( m_1^I n_2^I + m_1^J n_2^J \right) - \beta_D J_{21} N \left( m_1^I n_2^J + m_1^J n_2^I \right) +\nonumber\\
    &+ \alpha \tilde{\gamma} N \log2\cosh\left[ J_{1}\left( \beta_S n_1^I + \beta_D n_1^J \right) + J_{21}\left( \beta_S n_2^I + \beta_D n_2^J \right) \right] + \nonumber\\
    &+ (1-\alpha) \tilde{\gamma} N \log2\cosh\left[ J_{1}\left( \beta_S n_1^J + \beta_D n_1^I \right) + J_{21}\left( \beta_S n_2^J + \beta_D n_2^I \right) \right] +\nonumber\\
    &+ \alpha \tilde{\gamma} N \log2\cosh\left[ J_{1}\left( \beta_S m_1^I + \beta_D m_1^J \right) + J_{12}\left( \beta_S m_2^I + \beta_D m_2^J \right) \right] + \nonumber\\
    &+ (1-\alpha) \tilde{\gamma} N \log2\cosh\left[ J_{1}\left( \beta_S m_1^J + \beta_D m_1^I \right) + J_{12}\left( \beta_S m_2^J + \beta_D m_2^I \right) \right] + \nonumber\\
    &+ (\gamma - \alpha \tilde{\gamma}) N \log2\cosh\left[ J_{2}\left( \beta_S n_2^I + \beta_D n_2^J \right) + J_{12}\left( \beta_S n_1^I + \beta_D n_1^J \right) \right] + \nonumber\\
    &+ (1 - \gamma - (1-\alpha)\tilde{\gamma}) N \log2\cosh\left[ J_{2}\left( \beta_S n_2^J + \beta_D n_2^I \right) + J_{12}\left( \beta_S n_1^J + \beta_D n_1^I \right) \right] +\nonumber\\
    &+ (\gamma - \alpha \tilde{\gamma}) N \log2\cosh\left[ J_{2}\left( \beta_S m_2^I + \beta_D m_2^J \right) + J_{21}\left( \beta_S m_1^I + \beta_D m_1^J \right) \right] + \nonumber\\
    &+ (1 - \gamma - (1-\alpha)\tilde{\gamma}) N \log2\cosh\left[ J_{2}\left( \beta_S m_2^J + \beta_D m_2^I \right) + J_{21}\left( \beta_S m_1^J + \beta_D m_1^I \right) \right],
\end{align}
\normalsize
where we define $\alpha$ as the fraction of elements of the group 1 that belongs to the partition $I$. The fractional sizes for the other components can be obtained by imposing the constraint that the fractional size of the component $I$ must be $\gamma$ and the whole system must sum up to 1.

The saddle point equations are
\begin{eqnarray}
    m_1^I &=& \alpha \tilde{\gamma}\tanh\left[ J_{1}\left( \beta_S n_1^I + \beta_D n_1^J \right) + i J_{21}\left( \beta_S n_2^I + \beta_D n_2^J \right) \right],\\
    m_1^J &=& (1-\alpha) \tilde{\gamma}\tanh\left[ J_{1}\left( \beta_S n_1^J + \beta_D n_1^I \right) + i J_{21}\left( \beta_S n_2^J + \beta_D n_2^I \right) \right],\\
    m_2^I &=& \bar\alpha(1 - \tilde{\gamma}) \tanh\left[ J_{2}\left( \beta_S n_2^I + \beta_D n_2^J \right) + i J_{12}\left( \beta_S n_1^I + \beta_D n_1^J \right) \right], \\
    m_2^J &=& (1-\bar\alpha)\alpha(1 - \tilde{\gamma}) \tanh\left[ J_{2}\left( \beta_S n_2^J + \beta_D n_2^I \right) + i J_{12}\left( \beta_S n_1^J + \beta_D n_1^I \right) \right],\\
    n_1^I &=& \alpha \tilde{\gamma}\tanh\left[ J_{1}\left( \beta_S m_1^I + \beta_D m_1^J \right) + i J_{12}\left( \beta_S m_2^I + \beta_D m_2^J \right) \right],\\
    n_1^J &=& (1-\alpha) \tilde{\gamma}\tanh\left[ J_{1}\left( \beta_S m_1^J + \beta_D m_1^I \right) + i J_{12}\left( \beta_S m_2^J + \beta_D m_2^I \right) \right],\\
    n_2^I &=& \bar\alpha(1 - \tilde{\gamma}) \tanh\left[ J_{2}\left( \beta_S m_2^I + \beta_D m_2^J \right) + i J_{21}\left( \beta_S m_1^I + \beta_D m_1^J \right) \right],\\
    n_2^J &=& (1-\bar\alpha)(1 - \tilde{\gamma}) \tanh\left[ J_{2}\left( \beta_S m_2^J + \beta_D m_2^I \right) + i J_{21}\left( \beta_S m_1^J + \beta_D m_1^I \right) \right],
\end{eqnarray}
where we define
\begin{equation}
    \bar{\alpha} = \frac{\gamma - \alpha \tilde{\gamma}}{1 - \tilde{\gamma}},
\end{equation}
as the fraction of elements of the group 2 that belongs to the component $I$.

Taking the derivative
\begin{align}
    \frac{\partial}{\partial \beta_D} \log Z_A = N J_1 \left( m_1^I n_1^J + m_1^J n_1^I \right) + N J_{21} \left( m_1^I n_2^J + m_1^J n_2^I \right) + \nonumber\\
    + N J_{12} \left( m_2^I n_1^J + m_2^J n_1^I \right) + N J_2 \left( m_2^I n_2^J + m_2^J n_2^I \right),
\end{align}
where we used the equations of state to substitute the $\tanh$ for the correspondent order parameter.

Taking $\beta_S = \beta_D = \beta$ we have:
\begin{alignat}{2}
    &m_1^I = \alpha m_1, &&\qquad n_1^I = \alpha n_1,\\
    &m_1^J = (1-\alpha) m_1, &&\qquad n_1^J = (1-\alpha) n_1,\\
    &m_2^I = \bar{\alpha} m_2, &&\qquad n_2^I = \bar{\alpha} n_2,\\
    &m_2^J = (1-\bar{\alpha}) m_2, &&\qquad n_2^J = (1-\bar{\alpha}) n_2,\\
\end{alignat}
where $m_1$, $m_2$, $n_1$ and $n_2$ are now the order parameter of the modular Ising model.

And the value of $\phi_A$
\begin{align}
    \phi_A =& 2\alpha(1-\alpha) N \beta J_1 m_1 n_1 + \left[ \alpha(1-\bar{\alpha}) + \bar{\alpha}(1-\alpha) \right] N \beta J_{21} m_1 n_2 + \nonumber\\
    &+ \left[ \bar{\alpha}(1-\alpha) + \alpha(1-\bar{\alpha}) \right] N \beta J_{12} m_2 n_1 + 2\bar{\alpha}(1-\bar{\alpha}) N \beta J_2 m_2 n_2.
\end{align}

\subsubsection{$\log Z_B$ calculation}

\begin{equation}
    Z_B = \sum_X \exp\left\{ \lambda \sum_i \log2\cosh\left( \beta \sum_j J_{ij} x_j \right) \right\}.
\end{equation}

Introducing integrals over delta distributions $\delta\left( w_i - i\beta \sum_j J_{ij} x_j \right)$ and its Fourier representation
\small
\begin{equation}
    Z_B = \sum_X \int \prod_i \frac{dw_i d\hat{w}_i}{2\pi} \exp\left\{ \lambda \sum_i \log2\cosh\left( -i w_i \right) + i \sum_i \hat{w}_i w_i + \beta \sum_i \hat{w}_i \sum_j J_{ij} x_j \right\}.
\end{equation}
\normalsize

Introducing more integrals over delta distributions and using its Fourier representation for the order parameters:
\begin{align}
    Z_B =& \int \frac{dm_1 d\hat{m}_1}{2\pi/N} \int \frac{dm_2 d\hat{m}_2}{2\pi/N} \int \frac{dn_1 d\hat{n}_1}{2\pi/N} \int \frac{dn_2 d\hat{n}_2}{2\pi/N} \times \nonumber\\
    &\times \exp\bigg\{ \beta J_1 N m_1 n_1 + \beta J_{21} N m_1 n_2 + \beta J_{12} N m_2 n_1 + \beta J_2 N m_2 n_2 + \nonumber\\
    &+ i N \left( m_1 \hat{m}_1 + m_2 \hat{m}_2 + n_1 \hat{n}_1 + n_2 \hat{n}_2 \right) + \tilde{\gamma} N \log \mathcal{Z}_1 + (1 - \tilde{\gamma}) N \log \mathcal{Z}_2 \bigg\},
\end{align}
where
\begin{eqnarray}
    \mathcal{Z}_1 &=& 2\cosh(-i\hat{m}_1) \left[ 2\cosh(-i\hat{u}_1) \right]^\lambda, \\
    \mathcal{Z}_2 &=& 2\cosh(-i\hat{m}_2) \left[ 2\cosh(-i\hat{u}_2) \right]^\lambda.
\end{eqnarray}

Using the saddle point method of integration
\small
\begin{align}
    \log Z_B =& - \beta J_1 N m_1 n_1 - \beta J_{21} N m_1 n_2 - \beta J_{12} N m_2 n_1 - \beta J_2 N m_2 n_2 + \nonumber\\
    &+ \tilde{\gamma} N \log2\cosh[\beta \left( J_1 u_1 + J_{21} u_2 \right)] + \lambda \tilde{\gamma} N \log 2\cosh[\beta \left( J_1 m_1 + J_{21} m_2 \right)] + \nonumber\\
    &+ (1 - \tilde{\gamma}) N \log2\cosh[\beta \left( J_{12} u_1 + J_{2} u_2 \right)] + \lambda (1 - \tilde{\gamma}) N \log 2\cosh[\beta \left( J_{21} m_1 + J_{2} m_2 \right)].
\end{align}
\normalsize
The saddle point equations are:
\begin{eqnarray}
    m_1 &=& \tilde\gamma \tanh\left[ \beta \left( J_1 u_1 + J_{21} u_2 \right) \right],\\
    m_2 &=& (1-\tilde\gamma) \tanh\left[ \beta \left( J_{12} u_1 + J_{2} u_2 \right) \right],\\
    u_1 &=& \tilde\gamma \tanh\left[ \beta \left( J_1 m_1 + J_{21} m_2 \right) \right],\\
    u_2 &=& (1-\tilde\gamma) \tanh\left[ \beta \left( J_{21} m_1 + J_{2} m_2 \right) \right].
\end{eqnarray}

Taking the derivative with respect to $\lambda$, we obtain:
\begin{equation}
    \phi_B = - \tilde{\gamma} N \log 2\cosh\left[\beta \left( J_1 m_1 + J_{21} m_2 \right)\right] - (1 - \tilde{\gamma}) N \log 2\cosh\left[\beta \left( J_{21} m_1 + J_{2} m_2 \right)\right].
\end{equation}

\subsubsection{$\log Z_C$ calculation}

\begin{equation}
    Z_C = \sum_{X,Y} \exp\left\{ \beta\sum_{i,j} J_{ij} x_j y_i + \lambda_S \sum_i \log 2\cosh\left( \beta \sum_j J_{ij} \theta_{ij}^{S} x_j \right) \right\}.
\end{equation}

Again we introduce integrals over $\delta\left( w_i - i \beta \sum_j J_{ij} \theta_{ij}^{S} x_j \right)$, and use its Fourier representation
\begin{align}
    Z_C = \sum_{X,Y} \int \prod_i \frac{dw_i d\hat{w}_i}{2\pi} \exp\left\{ \beta\sum_{i,j} J_{ij} x_j y_i + \lambda_S \sum_i \log 2\cosh\left( -i w_i \right) +\right. \nonumber\\
    + \left.i\sum_i \hat{w}_i w_i + \beta \sum_i \hat{w}_i \sum_j J_{ij} \theta_{ij}^{S} x_j \right\},
\end{align}
where, again we used the projector $\theta_{ij}^S = \eta_i \eta_j + (1-\eta_i)(1-\eta_j)$ that selects all the interactions between the same component.

Introducing integrals over delta distribution for each parameter
\small
\begin{align}
    Z_C =& \int \frac{dm_1 d\hat{m}_1}{2\pi/N} \int \frac{dm_2 d\hat{m}_2}{2\pi/N} \int \frac{dn_1 d\hat{n}_1}{2\pi/N} \int \frac{dn_2 d\hat{n}_2}{2\pi/N} \int \frac{dm_1^I d\hat{m}_1^I}{2\pi/N} \int \frac{dm_1^J d\hat{m}_1^J}{2\pi/N} \times \nonumber\\
    &\times \int \frac{dm_2^I d\hat{m}_2^I}{2\pi/N} \int \frac{dm_2^J d\hat{m}_2^J}{2\pi/N} \int \frac{du_1^I d\hat{u}_1^I}{2\pi/N} \int \frac{du_1^J d\hat{u}_1^J}{2\pi/N}\int \frac{du_1^I d\hat{u}_1^I}{2\pi/N} \int \frac{du_1^J d\hat{u}_1^J}{2\pi/N} \times \nonumber\\
    &\times \exp \Bigg\{ \beta J_1 N m_1 n_1 + \beta J_{21} N m_1 n_2 + \beta J_{12} N m_2 n_1 + \beta J_2 N m_2 n_2 + \beta J_1 N \left( m_1^I u_1^I + m_1^J u_1^J \right) +\nonumber\\
    &+ \beta J_{21} N \left( m_1^I u_2^I + m_1^J u_2^J \right) + \beta J_{12} N \left( m_2^I u_1^I + m_2^J u_1^J \right) + \beta J_{2} N \left( m_2^I u_2^I + m_2^J u_2^J \right) +\nonumber\\
    &+ i N \big( m_1 \hat{m}_1 + m_2 \hat{m}_2 + n_1 \hat{n}_1 + n_2 \hat{n}_2 + m_1^I \hat{m}_1^I + m_1^J \hat{m}_1^J + m_2^I \hat{m}_2^I + m_2^J \hat{m}_2^J +\nonumber\\
    &+ u_1^I \hat{u}_1^I + u_1^J \hat{u}_1^J + u_2^J \hat{u}_2^J + u_2^J \hat{u}_2^J \big) + \sum_{i=1}^{M} \log \zeta_i^{(1)} + \sum_{i=M+1}^{N} \log \zeta_i^{(2)} \Bigg\},
\end{align}
\normalsize
where
\begin{align}
    \zeta_i^{(1)} =& 2\cosh\left(-i\hat{n}_1\right) \left[ 2\cosh \left(- i\left( \eta_i \hat{u}_1^I + (1-\eta_i)\hat{u}_1^J \right) \right)\right]^{\lambda_S} \times \nonumber\\
    &\times 2\cosh\left( -i\left( \hat{m}_1 + \eta_i \hat{m}_1^I + (1-\eta_i)\hat{m}_1^J \right) \right),
\end{align}
\begin{align}
    \zeta_i^{(2)} =& 2\cosh\left(-i\hat{n}_2\right) \left[ 2\cosh \left(- i\left( \eta_i \hat{u}_2^I + (1-\eta_i)\hat{u}_2^J \right) \right)\right]^{\lambda_S} \times \nonumber\\
    &\times 2\cosh\left( -i\left( \hat{m}_2 + \eta_i \hat{m}_2^I + (1-\eta_i)\hat{m}_2^J \right) \right).
\end{align}

The saddle point integration method gives us
\begin{align}
    \log Z_C =& \beta J_1 N m_1 n_1 + \beta J_{21} N m_1 n_2 + \beta J_{12} N m_2 n_1 + \beta J_2 N m_2 n_2 +\nonumber\\
    &+ \beta J_1 N \left( m_1^I u_1^I + m_1^J u_1^J \right) + \beta J_{21} N \left( m_1^I u_2^I + m_1^J u_2^J \right) +\nonumber\\
    &+ \beta J_{12} N \left( m_2^I u_1^I + m_2^J u_1^J \right) + \beta J_{2} N \left( m_2^I u_2^I + m_2^J u_2^J \right) +\nonumber\\
    &+ i N \big( m_1 \hat{m}_1 + m_2 \hat{m}_2 + n_1 \hat{n}_1 + n_2 \hat{n}_2 + m_1^I \hat{m}_1^I + m_1^J \hat{m}_1^J +\nonumber\\
    &+ m_2^I \hat{m}_2^I + m_2^J \hat{m}_2^J + u_1^I \hat{u}_1^I + u_1^J \hat{u}_1^J + u_2^J \hat{u}_2^J + u_2^J \hat{u}_2^J \big) + \nonumber\\
    &+ \tilde{\gamma} N \log 2\cosh\left( -i \hat{n}_1 \right) + (1-\tilde{\gamma}) N \log 2\cosh\left( -i \hat{n}_2 \right) + \nonumber\\
    &+ \sum_{i=1}^{M} \log 2\cosh\left( -i\left( \hat{m}_1 + \eta_i \hat{m}_1^I + (1-\eta_i)\hat{m}_1^J \right) \right) + \nonumber\\
    &+ \lambda_S \sum_{i=1}^{M} \log 2\cosh\left( - i\left( \eta_i \hat{u}_1^I + (1-\eta_i)\hat{u}_1^J \right) \right) + \nonumber\\
    &+ \sum_{i=M+1}^{N} \log 2\cosh\left( -i\left( \hat{m}_2 + \eta_i \hat{m}_2^I + (1-\eta_i)\hat{m}_2^J \right) \right) + \nonumber\\
    &+ \lambda_S \sum_{i=M+1}^{N} \log 2\cosh\left( - i\left( \eta_i \hat{u}_2^I + (1-\eta_i)\hat{u}_2^J \right) \right).
\end{align}
With the saddle point equations
\begin{alignat}{2}
    &\hat{m}_1 = i\beta \left( J_1 n_1 + J_{21} n_2 \right), &&\qquad \hat{n}_1 = i\beta \left( J_1 m_1 + J_{12} m_2 \right),\\
    &\hat{m}_2 = i\beta \left( J_{12} n_1 + J_{2} n_2 \right), &&\qquad \hat{n}_2 = i\beta \left( J_{21} m_1 + J_{2} m_2 \right),\\
    &\hat{m}_1^I = i\beta \left( J_1 u_1^I + J_{21} u_2^I \right), &&\qquad \hat{u}_1^I = i\beta \left( J_1 m_1^I + J_{12} m_2^I \right),\\
    &\hat{m}_1^J = i\beta \left( J_1 u_1^J + J_{21} u_2^J \right), &&\qquad \hat{u}_1^J = i\beta \left( J_1 m_1^J + J_{12} m_2^J \right),\\
    &\hat{m}_2^I = i\beta \left( J_{12} u_1^I + J_{2} u_2^I \right), &&\qquad \hat{u}_2^I = i\beta \left( J_{21} m_1^I + J_{2} m_2^I \right),\\
    &\hat{m}_2^J = i\beta \left( J_{12} u_1^J + J_{2} u_2^J \right), &&\qquad \hat{u}_2^J = i\beta \left( J_{21} m_1^J + J_{2} m_2^J \right),
\end{alignat}
\begin{eqnarray}
    n_1 &=& \tilde{\gamma} \tanh\left( -i\hat{n}_1 \right),\\
    n_2 &=& (1-\tilde{\gamma}) \tanh\left( -i\hat{n}_2 \right),\\
    m_1 &=& \sum_{i=1}^{M} \tanh\left[ -i\left( \hat{m}_1 + \eta_i \hat{m}_1^I + (1-\eta_i)\hat{m}_1^J \right) \right],\\
    m_1^I &=& \sum_{i=1}^{M} \eta_i \tanh\left[ -i\left( \hat{m}_1 + \eta_i \hat{m}_1^I + (1-\eta_i)\hat{m}_1^J \right) \right],\\
    m_1^J &=& \sum_{i=1}^{M} (1-\eta_i) \tanh\left[ -i\left( \hat{m}_1 + \eta_i \hat{m}_1^I + (1-\eta_i)\hat{m}_1^J \right) \right],\\
    m_2 &=& \sum_{i=M+1}^{N} \tanh\left[ -i\left( \hat{m}_2 + \eta_i \hat{m}_2^I + (1-\eta_i)\hat{m}_2^J \right) \right],\\
    m_2^I &=& \sum_{i=M+1}^{N} \eta_i \tanh\left[ -i\left( \hat{m}_2 + \eta_i \hat{m}_2^I + (1-\eta_i)\hat{m}_2^J \right) \right],\\
    m_2^J &=& \sum_{i=M+1}^{N} (1-\eta_i) \tanh\left[ -i\left( \hat{m}_2 + \eta_i \hat{m}_2^I + (1-\eta_i)\hat{m}_2^J \right) \right],\\
    u_1^I &=& \lambda_S\sum_{i=1}^{M} \eta_i \tanh\left[ -i\left( \eta_i \hat{u}_1^I + (1-\eta_i)\hat{u}_1^J \right) \right],\\
    u_1^J &=& \lambda_S \sum_{i=1}^{M} (1-\eta_i) \tanh\left[ -i\left( \eta_i \hat{u}_1^I + (1-\eta_i)\hat{u}_1^J \right) \right],\\
    u_2^I &=& \lambda_S \sum_{i=M+1}^{N} \eta_i \tanh\left[ -i\left( \eta_i \hat{u}_2^I + (1-\eta_i)\hat{u}_2^J \right) \right],\\
    u_2^J &=& \lambda_S \sum_{i=M+1}^{N} (1-\eta_i) \tanh\left[ -i\left( \eta_i \hat{u}_2^I + (1-\eta_i)\hat{u}_2^J \right) \right].
\end{eqnarray}

Now we are able to take the derivative of $\log Z_C$ with respect to $\lambda_S$:
\begin{align}
    \frac{\partial}{\partial \lambda_S}\log Z_C =& \sum_{i=1}^{M} \log 2\cosh\left( - i\left( \eta_i \hat{u}_1^I + (1-\eta_i)\hat{u}_1^J \right) \right) + \nonumber\\
    &+ \sum_{i=M+1}^{N} \log 2\cosh\left( - i\left( \eta_i \hat{u}_2^I + (1-\eta_i)\hat{u}_2^J \right) \right).
\end{align}

Taking $\lambda_S = 0$, we notice that the parameters $u_1^I$, $u_1^J$, $u_2^I$ and $u_2^J$ all are zero. Thus, in this case, the expression for the order parameters become
\begin{eqnarray}
    n_1 &=& \tilde{\gamma} \tanh\left( -i\hat{n}_1 \right),\\
    n_2 &=& (1-\tilde{\gamma}) \tanh\left( -i\hat{n}_2 \right),\\
    m_1 &=& \tilde{\gamma} \tanh\left( -i\hat{m}_1 \right),\\
    m_2 &=& (1-\tilde{\gamma}) \tanh\left( -i\hat{m}_2 \right),\\
    m_1^I &=& \sum_{i=1}^{M} \eta_i \tanh\left( -i\hat{m}_1 \right),\\
    m_1^J &=& \sum_{i=1}^{M} (1-\eta_i) \tanh\left( -i\hat{m}_1 \right),\\
    m_2^I &=& \sum_{i=M+1}^{N} \eta_i \tanh\left( -i\hat{m}_2 \right),\\
    m_2^J &=& \sum_{i=M+1}^{N} (1-\eta_i) \tanh\left( -i\hat{m}_2 \right),\\
\end{eqnarray}
and we notice that $m_1$, $m_2$, $n_1$ and $n_2$ are the order parameters of the full modular Ising model, and by using a similar notation we used before for $\phi_A$, we can write this order parameters in terms of same variable $\alpha$ that denotes the fraction of first $M$ elements that belongs to the component $I$:
\begin{eqnarray}
    m_1^I &=& \alpha m_1, \\
    m_1^J &=& (1-\alpha) m_1,\\
    m_2^I &=& \bar{\alpha} m_2, \\
    m_2^J &=& (1-\bar{\alpha}) m_2,
\end{eqnarray}
where $m_1$, $m_2$, $n_1$ and $n_2$ are the order parameters for the modular Ising model and, $\bar{\alpha}$ is, again, given by
\begin{equation}
    \bar{\alpha} = \frac{\gamma - \alpha \tilde\gamma}{1 - \tilde\gamma}.
\end{equation}

Finally we have:
\begin{align}
    \phi_C =& \alpha \tilde\gamma N \log 2\cosh\left[ \beta \left( J_1 \alpha m_1 + J_{12} \bar\alpha m_2 \right) \right] +\nonumber\\
    &+ (1 - \alpha) \tilde\gamma N \log 2\cosh\left[ \beta \left( J_1 (1-\alpha) m_1 + J_{12} (1-\bar\alpha) m_2 \right) \right] +\nonumber\\
    &+ \bar\alpha (1 - \tilde\gamma) N \log 2\cosh\left[ \beta \left( J_{21} \alpha m_1 + J_{2} \bar\alpha m_2 \right) \right] +\nonumber\\
    &+ (1 - \bar\alpha) (1 - \tilde\gamma) N \log 2\cosh\left[ \beta \left( J_{21} (1-\alpha) m_1 + J_{2} (1-\bar\alpha) m_2 \right) \right].
\end{align}
Defining the vectors
\begin{equation}
    \mathbf{m}_I = 
    \begin{pmatrix}
        \alpha m_1\\
        \bar{\alpha} m_2
    \end{pmatrix},
    \quad
    \mathbf{m}_J = 
    \begin{pmatrix}
        (1-\alpha) m_1\\
        (1-\bar{\alpha}) m_2
    \end{pmatrix},
    \quad
    \mathbf{n}_I = 
    \begin{pmatrix}
        \alpha n_1\\
        \bar{\alpha} n_2
    \end{pmatrix},
    \quad
    \mathbf{n}_J = 
    \begin{pmatrix}
        (1-\alpha) n_1\\
        (1-\bar{\alpha}) n_2
    \end{pmatrix},
\end{equation}
we obtain equation \eqref{phimodular}.

\end{document}